\documentclass[10pt,journal,compsoc]{IEEEtran}
\ifCLASSOPTIONcompsoc
  \usepackage[nocompress]{cite}
\else
  \usepackage{cite}
\fi
\ifCLASSINFOpdf
  \usepackage[pdftex]{graphicx}
  \DeclareGraphicsExtensions{.pdf,.jpeg,.png}
\else
  \usepackage[dvips]{graphicx}
  \DeclareGraphicsExtensions{.eps}
\fi
\usepackage{amsmath}
\ifCLASSOPTIONcompsoc
 \usepackage[caption=true,font=footnotesize,labelfont=sf,textfont=sf]{subfig}
\else
 \usepackage[caption=true,font=footnotesize]{subfig}
\fi

\usepackage{stfloats}
\usepackage{url}
\usepackage{amsfonts}
\usepackage{amssymb}
\usepackage{bm}
\usepackage{color}
\usepackage[ruled, linesnumbered]{algorithm2e}
\usepackage{tikz}
\usepackage{caption}
\usepackage{multirow}
\usepackage{blindtext}
\usepackage{adjustbox}

\newtheorem{definition}{Definition}
\newtheorem{lemma}{Lemma}
\newtheorem{theorem}{Theorem}

\DeclareMathOperator*{\argmin}{argmin}

\definecolor{myGrey}{RGB}{192,192,192}
\definecolor{ForestGreen}{rgb}{0.13,0.55,0.13}
\definecolor{NavyBlue}{RGB}{0,102,204}
\definecolor{myBrown}{RGB}{145,60,0}

\begin{document}
%
% paper title
% Titles are generally capitalized except for words such as a, an, and, as,
% at, but, by, for, in, nor, of, on, or, the, to and up, which are usually
% not capitalized unless they are the first or last word of the title.
% Linebreaks \\ can be used within to get better formatting as desired.
% Do not put math or special symbols in the title.
\title{Differentially Private $k$-Means Clustering with Guaranteed Convergence}
%
%
% author names and IEEE memberships
% note positions of commas and nonbreaking spaces ( ~ ) LaTeX will not break
% a structure at a ~ so this keeps an author's name from being broken across
% two lines.
% use \thanks{} to gain access to the first footnote area
% a separate \thanks must be used for each paragraph as LaTeX2e's \thanks
% was not built to handle multiple paragraphs
%
%
%\IEEEcompsocitemizethanks is a special \thanks that produces the bulleted
% lists the Computer Society journals use for "first footnote" author
% affiliations. Use \IEEEcompsocthanksitem which works much like \item
% for each affiliation group. When not in compsoc mode,
% \IEEEcompsocitemizethanks becomes like \thanks and
% \IEEEcompsocthanksitem becomes a line break with idention. This
% facilitates dual compilation, although admittedly the differences in the
% desired content of \author between the different types of papers makes a
% one-size-fits-all approach a daunting prospect. For instance, compsoc 
% journal papers have the author affiliations above the "Manuscript
% received ..."  text while in non-compsoc journals this is reversed. Sigh.

\author{Zhigang~Lu,
        and Hong~Shen,% <-this % stops a space
\IEEEcompsocitemizethanks{\IEEEcompsocthanksitem Z. Lu and H. Shen are with School
of Computer Science, the University of Adelaide, Adelaide,
SA, 5005, Australia.\protect\\
% note need leading \protect in front of \\ to get a newline within \thanks as
% \\ is fragile and will error, could use \hfil\break instead.
E-mail: \{zhigang.lu, hong.shen\}@adelaide.edu.au
\IEEEcompsocthanksitem H. Shen is with School of Data and Computer Science, Sun Yat-sen University, Guangzhou, Guangdong, 510006, China.}% <-this % stops an unwanted space
\thanks{Manuscript received Month Date, Year; revised Month Date, Year.}}

% note the % following the last \IEEEmembership and also \thanks - 
% these prevent an unwanted space from occurring between the last author name
% and the end of the author line. i.e., if you had this:
% 
% \author{....lastname \thanks{...} \thanks{...} }
%                     ^------------^------------^----Do not want these spaces!
%
% a space would be appended to the last name and could cause every name on that
% line to be shifted left slightly. This is one of those "LaTeX things". For
% instance, "\textbf{A} \textbf{B}" will typeset as "A B" not "AB". To get
% "AB" then you have to do: "\textbf{A}\textbf{B}"
% \thanks is no different in this regard, so shield the last } of each \thanks
% that ends a line with a % and do not let a space in before the next \thanks.
% Spaces after \IEEEmembership other than the last one are OK (and needed) as
% you are supposed to have spaces between the names. For what it is worth,
% this is a minor point as most people would not even notice if the said evil
% space somehow managed to creep in.

% The paper headers
\markboth{IEEE Transactions On ABC,~Vol.~X, No.~Y, Month~Year}%
{Lu \MakeLowercase{\textit{et al.}}: Differentially Private $k$-Means Clustering with Guaranteed Convergence}
% The only time the second header will appear is for the odd numbered pages
% after the title page when using the twoside option.
% 
% *** Note that you probably will NOT want to include the author's ***
% *** name in the headers of peer review papers.                   ***
% You can use \ifCLASSOPTIONpeerreview for conditional compilation here if
% you desire.

% The publisher's ID mark at the bottom of the page is less important with
% Computer Society journal papers as those publications place the marks
% outside of the main text columns and, therefore, unlike regular IEEE
% journals, the available text space is not reduced by their presence.
% If you want to put a publisher's ID mark on the page you can do it like
% this:
%\IEEEpubid{0000--0000/00\$00.00~\copyright~2015 IEEE}
% or like this to get the Computer Society new two part style.
%\IEEEpubid{\makebox[\columnwidth]{\hfill 0000--0000/00/\$00.00~\copyright~2015 IEEE}%
%\hspace{\columnsep}\makebox[\columnwidth]{Published by the IEEE Computer Society\hfill}}
% Remember, if you use this you must call \IEEEpubidadjcol in the second
% column for its text to clear the IEEEpubid mark (Computer Society jorunal
% papers don't need this extra clearance.)

% use for special paper notices
%\IEEEspecialpapernotice{(Invited Paper)}

% for Computer Society papers, we must declare the abstract and index terms
% PRIOR to the title within the \IEEEtitleabstractindextext IEEEtran
% command as these need to go into the title area created by \maketitle.
% As a general rule, do not put math, special symbols or citations
% in the abstract or keywords.
\IEEEtitleabstractindextext{%
\begin{abstract}
Iterative clustering around representative points as an effective technique for clustering helps us learn the insights behind data and enables various important applications to build on. Unfortunately, it also provides security holes which may allow adversaries to infer the privacy of individuals with some background knowledge. To protect individual privacy against such inference attacks, preserving differential privacy for iterative clustering algorithms has been extensively studied. Existing differentially private clustering algorithms adopt the same framework to compute differentially private centroids iteratively: running Lloyd's $k$-means algorithm to obtain the real centroids, then perturbing them with a differential privacy mechanism. These algorithms suffer from the non-convergence problem, i.e., they provide no guarantee of terminate at a solution of Lloyd's algorithm within a bounded number of iterations. This problem severely impacts their  clustering quality and execution efficiency. 

To address this problem, in this paper, following the same centroids updating pattern as existing work in the interactive setting, we propose a novel framework for injecting differential privacy into the real centroids in the interactive setting. Specifically, to ensure the convergence, we maintain the perturbed centroids of the previous iteration  $t-1$ to compute a \textit{convergence zone} for each cluster in the current iteration $t$, where we inject differential privacy noise. To have a satisfactory convergence rate, we further control the orientation of centroid movement in each cluster by two strategies: one takes the orientation of centroid movement from iteration $t-1$ to iteration $t$ (past knowledge); the other uses the additional information of the orientation from iteration $t$ to iteration $t+1$ (future knowledge). We prove that, in the expected case, our algorithm (in both strategies) converges to a solution of Lloyd's algorithm in at most twice as many iterations as Lloyd's algorithm. Furthermore, when using both past and future knowledge, we prove that our algorithm converges to the same solution as Lloyd's algorithm (for the same initial centroids) with high probability, at the cost of a slower convergence speed than using only past knowledge due to duplicated operations in each iteration required for computing the future knowledge. We perform experimental evaluations on six widely used real-world datasets. The experimental results show that our algorithm outperforms the state-of-the-art methods of interactive differentially private clustering with a guaranteed convergence and better clustering quality to meet the same differential privacy requirement.
\end{abstract}

% Note that keywords are not normally used for peerreview papers.
}

% make the title area
\maketitle

% To allow for easy dual compilation without having to reenter the
% abstract/keywords data, the \IEEEtitleabstractindextext text will
% not be used in maketitle, but will appear (i.e., to be "transported")
% here as \IEEEdisplaynontitleabstractindextext when the compsoc 
% or transmag modes are not selected <OR> if conference mode is selected 
% - because all conference papers position the abstract like regular
% papers do.
\IEEEdisplaynontitleabstractindextext
% \IEEEdisplaynontitleabstractindextext has no effect when using
% compsoc or transmag under a non-conference mode.

% For peer review papers, you can put extra information on the cover
% page as needed:
% \ifCLASSOPTIONpeerreview
% \begin{center} \bfseries EDICS Category: 3-BBND \end{center}
% \fi
%
% For peerreview papers, this IEEEtran command inserts a page break and
% creates the second title. It will be ignored for other modes.
\IEEEpeerreviewmaketitle

\IEEEraisesectionheading{\section{Introduction}\label{sec:introduction}}
% Computer Society journal (but not conference!) papers do something unusual
% with the very first section heading (almost always called "Introduction").
% They place it ABOVE the main text! IEEEtran.cls does not automatically do
% this for you, but you can achieve this effect with the provided
% \IEEEraisesectionheading{} command. Note the need to keep any \label that
% is to refer to the section immediately after \section in the above as
% \IEEEraisesectionheading puts \section within a raised box.

% The very first letter is a 2 line initial drop letter followed
% by the rest of the first word in caps (small caps for compsoc).
% 
% form to use if the first word consists of a single letter:
% \IEEEPARstart{A}{demo} file is ....
% 
% form to use if you need the single drop letter followed by
% normal text (unknown if ever used by the IEEE):
% \IEEEPARstart{A}{}demo file is ....
% 
% Some journals put the first two words in caps:
% \IEEEPARstart{T}{his demo} file is ....
% 
% Here we have the typical use of a "T" for an initial drop letter
% and "HIS" in caps to complete the first word.
\IEEEPARstart{I}{n} the era of big data analytics, along with the rapid development of deep learning and its impressive achievements, e.g., the Google AI Go player Alpha Go beats the best human Go player by self-taught with the deep neural networks~\cite{SilverD2016,WIKI2018}, traditional machine learning techniques, such as the $k$-means clustering algorithm, shows increasing importance for learning insights from the ``small data" without the ground truth, due to their attractiveness of high running efficiency and prediction accuracy~\cite{QARDAJI2013,NguyenT2016}. In this paper, we address the issue of effective privacy-preserving realization of the popular  Lloyd's $k$-means clustering~\cite{LLOYD1982} algorithm.

\begin{figure*}[!th]
\centering
\begin{tikzpicture}[xscale=0.45, yscale=0.45]
\node [above] at (0,5) {Cluster $i$ at Iteration $t$:};
\draw (0,0) circle [radius=5];
\draw [fill=myGrey] (2.5,3.5) -- (3,2.5) -- (2,2.5) -- (2.5,3.5);
\node [below] at (2.5,2.5) {$x_{0}$};
\draw [fill=black] (-1.5,4) -- (-2,3) -- (-1,3) -- (-1.5,4);
\draw [fill=black] (-3,3) -- (-2.5,2) -- (-3.5,2) -- (-3,3);
\draw [fill=black] (0,2) -- (0.5,1) -- (-0.5,1) -- (0,2);
\draw [fill=black] (-3,1) -- (-2.5,0) -- (-3.5,0) -- (-3,1);
\draw [fill=black] (1.5,0) -- (2,-1) -- (1,-1) -- (1.5,0);
\draw [fill=black] (0.5,-1) -- (1,-2) -- (0,-2) -- (0.5,-1);
\draw [fill=black] (-3.5,-1) -- (-3,-2) -- (-4,-2) -- (-3.5,-1);
\draw [fill=black] (-2.5,-2) -- (-2,-3) -- (-3,-3) -- (-2.5,-2);
\draw [fill=black] (-0.5,-3) -- (0,-4) -- (-1,-4) -- (-0.5,-3);
\draw [fill=ForestGreen] (-0.3,0) circle [radius=0.35];
\draw [fill=red] (-1.5,-0.8) circle [radius=0.35];
\node [above, draw] at (12,-3) {\textcolor{ForestGreen}{$S^{(t)}_{i}$}: Cluster centroid (include $x_{0}$)};
\draw [dashed, ForestGreen, <-] (0.05,0) to [out=0, in=180] (4.9,-2.3);
\node [above, draw] at (11.3,-5) {\textcolor{red}{$S^{\prime(t)}_{i}$}: Cluster centroid (exclude $x_{0}$)};
\draw [dashed, red, <-] (-1.15,-0.8) to [out=0, in=180] (4,-4.3);
\node [align=left, draw, dashed, thick] at (12,1.8) {The adversary knows:\\\textcolor{ForestGreen}{$S^{(t)}_{i}$}, \textcolor{red}{$S^{\prime(t)}_{i}$}, and $N^{(t)}_{i}$\\The adversary infers:\\$x_{0} = N^{(t)}_{i}\textcolor{ForestGreen}{S^{(t)}_{i}} - (N^{(t)}_{i} - 1)\textcolor{red}{S^{\prime(t)}_{i}}$};
\end{tikzpicture}
\caption{An Illustrated Example of The Inference Attack.}
\label{fig:7:attack}
\end{figure*}
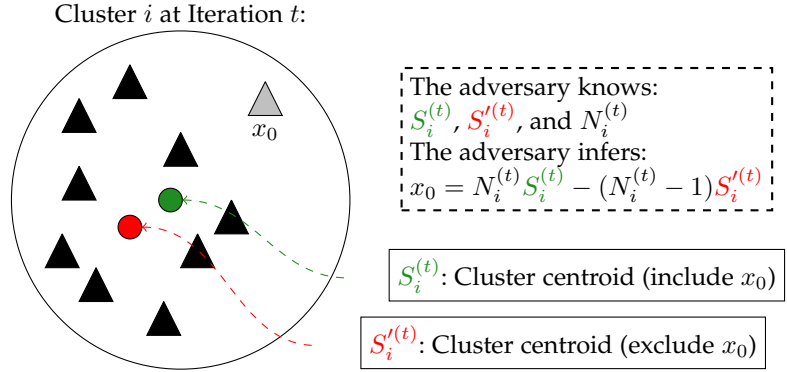

Despite the benefits we enjoyed from clustering, the privacy disclosure risk thwarts people's willingness to contribute data (especially the data that may link to privacy) to the clustering algorithms. Consider the following inference attack by the difference between the outputs from a private dataset and an adversary's background knowledge. There are a trusted data curator who manages a dataset $X$ and an adversary who owns a dataset $X'$. In the worst case, we have $\{x_0\} = X - X'$. At any arbitrary iteration $t$ of clustering, assume a set of centroids in $X$ is accidentally disclosed to the adversary. By comparing the difference between the set of centroids generated by $X$ and $X'$, the adversary can easily infer the value of the missing item $x_0$, thus technically gains the full access to the dataset $X$. Figure~\ref{fig:7:attack} depicts how such an inference attack works, where $n^{(t)}_{i}$ is the overall number of items in cluster $i$ at iteration $t$ of $X$.

From the above  inference attack example, it is clear that preserving the privacy of individual items in a dataset when running an iterative clustering algorithm needs to protect the true value of the centroids of the clusters at each iteration. Unfortunately, some of the well-known privacy preserving paradigms, such as Secure Multi-party Computation (SMC or MPC)~\cite{YAO1982} and Anonymity~\cite{SWEENEY2002,MACHANAVAJJHALA2007,LI2007}, are vulnerable to such an inference attack because both the SMC paradigm and the family of anonymity are vulnerable against the adversaries who have the maximum background knowledge (e.g., $n-1$ out of $n$ items of a dataset).

To preserve privacy against the inference attacks with maximum background knowledge (Figure~\ref{fig:7:attack}), differential privacy (DP)~\cite{DWORK2006B} has been applied in Lloyd's algorithm in the interactive setting~\cite{SU2017} whereby random DP noises were injected into each iteration when running Lloyd's algorithm. In a nutshell, there are a long line of studies~\cite{BLUM2005,DWORK2011,MOHAN2012,ZHANGJ2013,SU2017,PARK2017} guarantee DP while achieving acceptable clustering quality in the interactive setting via three DP mechanisms: the sample and aggregation framework of DP~\cite{NISSIM2007}, the exponential mechanism of DP (ExpDP)~\cite{MCSHERRY2007}, and the Laplace mechanism of DP (LapDP)~\cite{DWORK2006A}. We observed two weaknesses from existing work ~\cite{BLUM2005,DWORK2011,MOHAN2012,ZHANGJ2013,SU2017,PARK2017}. Particularly, the work~\cite{MOHAN2012}, with the sample and aggregation framework, showed unsatisfactory clustering quality because its uniform sampling may result skewness over the sampled buckets then the aggregated centroids would have a significant distance to the Lloyd's result. The studies~\cite{BLUM2005,DWORK2011,ZHANGJ2013,SU2017,PARK2017} applied ExpDP and LapDP suffered from a non-convergence problem since the unbounded noises are injected to an arbitrary direction. \textcolor{black}{The necessity of the convergence (defined in Definition~\ref{def:7:conv}) guarantee is two-fold. First, without convergence guarantee, a predefined iteration number is required to terminate a differentially private $k$-means algorithm. To find such an iteration number to satisfy the clustering quality with a given input dataset, we have to run the algorithm over the dataset multiple times. Furthermore, deploying the algorithm to different datasets needs to re-calculate the iteration number with the above process repeatedly which results in a large computational cost for this predefined parameter. Second, the non-convergent result may have a large distance to one of the local optimal solutions of the $k$-means problem, the clustering quality of the existing work~\cite{BLUM2005,DWORK2011,ZHANGJ2013,SU2017,PARK2017} is not always guaranteed. Therefore, this non-convergence problem severely impacts the efficiency and the clustering quality of applying the algorithm in the real life.}

To overcome the above weaknesses, we propose a new differentially private $k$-means clustering algorithm in the interactive setting that improves the existing work with a guaranteed convergence \textcolor{black}{(defined in Definition~\ref{def:7:conv})} and better clustering quality on the same DP requirement. In summary, our main contributions are:
\begin{itemize}
  \setlength\itemsep{0em}
  %\item To the best of our knowledge, this is the first work to explore the convergence of a differentially private $k$-means clustering algorithm in the interactive setting.
  \item We propose a novel approach of differentially private clustering that injects bounded DP noise into each iteration of the clustering process by applying ExpDP in a controlled orientation of progressing to preserve data privacy against inference attacks. In comparison to existing work which injects unbounded noise to arbitrary direction, our approach ensures convergence in at most doubled number of iterations as the Lloyd's $k$-means clustering.
  \item We mathematically analyse the key properties (convergence, the convergence rate, and the bound of DP) of our differentially private $k$-means clustering  algorithm for two centroids updating strategies, respectively,  based on past knowledge of previous-iteration centroids movement (same assumptions as existing work), and past and future knowledge --- centroids movements of previous and next iterations. The former requires fewer iterations for convergence, while the latter results in a better convergence quality.
  \item We experimentally evaluate the performance of clustering quality across various experimental settings on six widely used real-world datasets. With the same DP guarantee (privacy), because of the convergence guarantee, our algorithm for both centroids updating strategies  achieves better clustering quality (utility) than the state-of-the-art differentially private $k$-means clustering algorithms.
\end{itemize}

To the best of our knowledge, our algorithm is the first one that ensures convergence for differentially private $k$-means clustering in the interactive setting. 

% In comparison with our preliminary version~\cite{LuZ2019} of this paper, we propose a new centroids updating strategy in our differentially private $k$-means clustering. Specifically, in our preliminary version, we use the future knowledge of centroids movement, i.e., centroid position of next iteration, to navigate the centroid updating, which guarantees converging to the same centroids as the Lloyd's algorithm with high probability. However, due to the extra computations for the future centroid position in current iteration, the convergence rate is not satisfying in practice. Therefore, in this paper, to accelerate the convergence of our approach in practice, we introduce a novel way to update the centroids in each iteration, which does not rely on the future knowledge, but an estimation of the future centroids movement.

The rest of this paper is organised as follows: In Section~\ref{sec:7:relatedWork}, we discuss existing work on differentially private clustering in the interactive setting for both the advantages and disadvantages. In Section~\ref{sec:7:preliminaries}, we give a brief introduction of the preliminaries of this paper, including Lloyd's algorithm and DP. In Section~\ref{sec:7:dpkm} we introduce our approach to ensure convergence through  noise injection in controlled centroids movement orientation and preliminary analysis on the convergence property. In Section~\ref{sec:7:samplingzone}, we propose two designs of noise sampling zone in each iteration of clustering. In Section~\ref{sec:7:algorithm}, we describe our differentially private $k$-means clustering algorithm and its convergence and differential privacy proof. In Section~\ref{sec:7:experiments}, we provide the experimental evaluation to compare the clustering quality (data utility) of existing work and our algorithm. Finally, we conclude this paper in Section~\ref{sec:7:conclusion}.

\section{Related Work}
\label{sec:7:relatedWork}
In this section, we briefly summarise the related work on differentially private $k$-means clustering~\cite{BLUM2005,DWORK2011,LEI2011,MOHAN2012,ZHANGJ2013,SU2016,SU2017,PARK2017} in the interactive setting. In general, the results in the interactive setting with a DP guarantee deployed three major mechanisms of DP: the Laplace mechanism (LapDP)~\cite{DWORK2006A}, the sample and aggregation framework~\cite{NISSIM2007}, and the exponential mechanism (ExpDP)~\cite{MCSHERRY2007}.

There is a group of studies~\cite{BLUM2005,DWORK2011,SU2017} injected Laplace noise to the iterations of Lloyd's algorithm directly to ensure DP. The difference among these studies is the way to allocate privacy budget to each iteration. Blum et al.~\cite{BLUM2005} split the overall privacy budget uniformly to each iteration, prior to that, a total number of iterations was determined empirically. In spite of its simplicity, this scheme requires significant computational resources as it has to repeatedly run the algorithm on the target dataset to have a suitable number of iterations. Su et al.~\cite{SU2017} improved the weaknesses of \cite{BLUM2005} by allocating the privacy budget with a theoretically guaranteed optimal allocation method. However, this optimal allocation scheme may not fit all real-world datasets, as it assumes that all the clusters always have the same size. Dwork~\cite{DWORK2011} allocated the privacy budget with a decreasing exponential distribution, that is, assigned $1/2^{i}$ of the overall privacy budget at iteration $i$ until using up the overall privacy budget. Unfortunately, this scheme results in unsatisfactory clustering quality since the injected noises keep increasing when the allocated privacy budget is decreasing.

The sample and aggregation framework and the ExpDP were also used to ensure DP for an interactive $k$-means clustering algorithm. Mohan et al.~\cite{MOHAN2012} proposed GUPT applied the sample and aggregation framework of DP with Lloyd's algorithm. Briefly, GUPT uniformly samples items from an input dataset to different buckets, where local clustering result of each bucket is generated by Lloyd's algorithm. The final clustering result is the mean of those local ones with Laplace noise. Although GUPT is convergent, the clustering quality is unsatisfying because its uniform sampling may sample items from one cluster to a bucket with high probability, then the clustering result in such bucket will contribute a large amount noise to the aggregation stage. Zhang et al.~\cite{ZHANGJ2013} proposed a genetic algorithm (GA) based differentially private $k$-means clustering algorithm, PrivGene. Unlike the traditional GA, PrivGene randomly sampled the candidates for the next iteration with the ExpDP rather than selecting the top-quality ones. PrivGene achieves fair clustering quality if the input dataset is relatively small because in this case, it produces global optimal clustering result with high probability. However, similar to \cite{BLUM2005}, PrivGene also requires a predefined iteration number to terminate the algorithm. So efficiency would be a major problem to it. Differing from the above algorithms, Park et al.~\cite{PARK2017} achieved $(\epsilon, \delta)$-DP, rather than $\epsilon$-DP, with given assumption on the distribution of the input dataset which narrows its applicability in the real-world scenarios.

Based on the above analysis of the existing differentially private $k$-means clustering algorithms, we conclude that the convergence is an important property to the clustering quality of an iterative $k$-means clustering algorithm. Furthermore, it is essential to have a good trade-off between the privacy of each single item in a dataset and the clustering quality. Therefore, in this paper, we aim to explore how to guarantee convergence and better clustering quality to meet the same DP requirement as existing work in the interactive setting.

\section{Preliminaries}
%privacy notion
\label{sec:7:preliminaries}
In this section, we briefly introduce the notion of privacy used in this paper, i.e., differential privacy and Lloyd's $k$-means clustering algorithm. Following the same pattern as the existing differentially private $k$-means algorithms, the differential privacy noise is injected to the real centroids computed by Lloyd's algorithm over iterations.

\subsection{Differential Privacy.} Informally, DP is a scheme that minimises the sensitivity of output for a given statistical operation on two neighbouring (differentiated in one arbitrary record to protect) datasets. That is, DP guarantees the presence or absence of any item in a dataset will be concealed to the adversary with maximum auxiliary information.

In DP, the basic setting is a pair of neighbouring datasets $X$ and $X^{\prime}$, where $X^{\prime}$ contains the information of all the items except one item in a dataset $X$. A formal definition of Differential Privacy is shown as follow:
\begin{definition}[$\epsilon$-DP~\cite{DWORK2006A}]
\label{def:7:dp}
A randomised mechanism $\mathcal{T}$ is $\epsilon$-differentially private if for all neighbouring datasets $X$ and $X^{\prime}$, and for an arbitrary answer $s \in Range(\mathcal{T})$, $\mathcal{T}$ satisfies: 
$$\Pr[\mathcal{T}(X) = s]\leq \exp(\epsilon)\cdot\Pr[\mathcal{T}(X^{\prime}) = s]$$
where $\epsilon$ is the privacy budget.
\end{definition}

Two parameters are essential to DP: the privacy budget $\epsilon$ and the local function sensitivity $\Delta f$, i.e. $\Delta f(X)$, where $f$ is the query function to the dataset $X$. The privacy budget $\epsilon$ is set by the trusted dataset curator (who has full access to dataset $X$). Theoretically, a smaller $\epsilon$ denotes a higher privacy guarantee because the privacy budget $\epsilon$ reflects the magnitude of the difference between two neighbouring datasets. The reason why we use the local sensitivity is that it offers better utility to respond query $f$ when guaranteeing $\epsilon$-DP. $\Delta f$ is calculated by the following equation,
\begin{equation}
\label{eq:7:ls}
\Delta f(X) = \max_{\forall X'}|f(X) - f(X')|,
\end{equation}

In this paper, we mainly use two main mechanisms of DP: the Laplace mechanism (LapDP)~\cite{DWORK2006B} and the Exponential mechanism (ExpDP)~\cite{MCSHERRY2007}. In general, the LapDP adds random noise with Laplace distribution for the numeric computation to satisfy Definition~\ref{def:7:dp}. While for the non-numeric computation, the ExpDP introduces a scoring function $q(X, x)$ which reflects how appealing the pair $(X, x)$ is, where $X$ denotes a dataset and $x$ is the random respond to a query function on the dataset $X$. When applying the ExpDP, we can simply treat it as a weighted sampling, where the scoring function assigns weights to the sample space.In this paper, we mainly use two main mechanisms of DP: The formal definition is shown below:
\begin{definition}[Exponential Mechanism \cite{MCSHERRY2007}]
\label{def:7:expMech}
Given a scoring function of a dataset $X$, $q(X,x)$, which reflects the quality of query respond $x$. The exponential mechanism $\mathcal{T}$ provides $\epsilon$-differential privacy, if $\mathcal{T}(X) = \left\{ \Pr[x] \propto \exp\left({\frac{\epsilon\cdot q(X,x)}{2\Delta q}}\right) \right\}$, where $\Delta{q}$ is the sensitivity of scoring function $q(X, x)$, $\epsilon$ is the privacy budget.
\end{definition}

\subsection{Lloyd's $k$-Means Algorithm.}
The $k$-means clustering aims to split a dataset with $N$ items into $k$ clusters where each item is allocated into a cluster with the nearest cluster centroid to itself. The formal cost function of $k$-means clustering is:
\begin{equation}
\label{eq:7:kmeans}
\argmin_{\mathbf{C}} J = \sum_{i = 1}^{k}\sum_{x \in C_{i}}||x - S_{i}||^{2},
\end{equation}
where $\mathbf{C} = \{C_{1}, C_{2}, \dots, C_{k}\}$ is the set of $k$ clusters, $x$ is an item in the dataset $X = \{x_{1}, x_{2}, \dots, x_{N}\}$, $S_{i}$ is the centroid of $C_{i}$. Equation~\ref{eq:7:kmeans} calculates the total cost of a set of centroids.

The most well known $k$-means clustering algorithm is an iterative refinement algorithm called Lloyd's $k$-means clustering algorithm~\cite{LLOYD1982}. In brief, Lloyd's algorithm improves the quality of centroids by iteratively running a \textit{re-assignment} step and a \textit{re-centroid} step. In the \textit{re-assignment} step, it assigns each item to its nearest centroid to build the $k$ clusters. In the \textit{re-centroid} step, it re-calculates the centroid (mean) for each cluster. This new/updated $k$ centroids are used for the next \textit{re-assignment} step. Lloyd's algorithm terminates itself when the $k$ centroids keep the same in two neighbouring iterations. \textcolor{black}{Namely, Lloyd's algorithm is guaranteed to converge to one of the local optimal solutions of the $k$-means problem within finite iterations.}

Finally, to measure the quality of convergence, in this paper, we define \textit{convergence} and \textit{convergent degree} for a differentially private $k$-means clustering algorithm.
\begin{definition}[Convergence]
\label{def:7:conv}
Given a dataset $\mathcal{D}$, \textcolor{black}{an integer $k$,} Lloyd's algorithm $\mathcal{L}$, \textcolor{black}{and the set of local optimal solutions of the $k$-means problem $\mathcal{C}$,} we have $\mathcal{L}(\mathcal{D}) \rightarrow \mathcal{C}$. We say a differentially private $k$-means algorithm, $\mathcal{F}$, is convergent, i.f.f., $\mathcal{F}(\mathcal{D}) \rightarrow \mathcal{C}$.
\end{definition}

\begin{definition}[Convergence Degree]
Given a set of initial centroids $d \in \mathcal{D}$ and $\mathcal{L}(d) \rightarrow c \in \mathcal{C}$, the convergence degree of $\mathcal{F}$ is the probability $\Pr[\mathcal{F}(d) \rightarrow c]$.
\end{definition}

In addition, Table~\ref{tab:7:terms} lists the notations used in this paper.
\begin{table*}[!th]
\caption{The summary of notations.}
\label{tab:7:terms}
\centering
\begin{tabular}{c|l}
\hline
Notation & Description\\
\hline
\hline
$\hat{\cdot}$ & The corresponding notation ($\cdot$ from Lloyd's algorithm) in privacy-preserving algorithms\\
\hline
$a^{(t)}_{i}$ & Distance between $S^{(t)}_{i}$ and $\hat{S}^{(t)}_{i}$\\
\hline
$b^{(t)}_{i}$ & Distance between $S^{(t)}_{i}$ and $S^{(t+1)}_{i}$\\
\hline
$C^{(t)}_{i}$ & Cluster $i$ at iteration $t$\\
\hline
$\Delta$ & Value difference of the cost function between two iterations\\
\hline
$\epsilon^{(t)}_{i}$ & Differential privacy budget for $C^{(t)}_{i}$\\
\hline
$I$ & Overall iterations of Lloyd's algorithm\\
\hline
$q$ & Quality function from differential privacy\\
\hline
$J^{(S^{(t)}_{i})}$ & Value of the cost function for $C^{(t)}_{i}$ with centroid $S^{(t)}_{i}$\\
\hline
$S^{(t)}_{i}$ & Cluster centroid in $C^{(t)}_{i}$\\
\hline
\end{tabular}
\end{table*}

%Main
\section{Noise Injection in Controlled Orientation }
\label{sec:7:dpkm}
In this section, we first provide an overview of our approach, then preliminarily analyse the convergence property for a randomised centroids updating for $k$-means clustering.

% Overview
\subsection{Approach Overview}
\label{sec:7:overview}
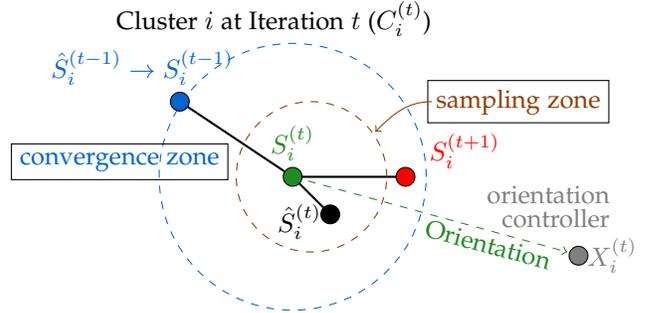
\begin{figure}[!th]
\centering
\begin{tikzpicture}[xscale=0.5, yscale=0.5]
% title
\node [above] at (2.5,1.5) {Cluster $i$ at Iteration $t$ ($C^{(t)}_{i}$)};
% convergence zone
\draw [dashed, NavyBlue] (3, -2) circle [radius=3.55];
% link the nodes
\draw [thick] (0,0) -- (3,-2) -- (6,-2);
% \draw [dashed] (6,-2) -- (14,-5);
% orientation
\draw [ForestGreen, dashed, ->] (3,-2) -- (10.3,-4);
% orientation label
\node [rotate=345] at (8.2,-3.8) {\textcolor{ForestGreen}{Orientation}};
% link to hat
\draw [thick] (3,-2) -- (4,-3);
% sampling zone
\draw [dashed, myBrown] (3.5,-2) circle [radius=2];
% sampling zone label
\node [draw] at (9,0) {\textcolor{myBrown}{sampling zone}};
% \node [below] at (12,-0.5) {\textcolor{myBrown}{(defined in Lemma~\ref{thm:7:sameconv})}};
\draw [myBrown, ->] (6.7,0) to [out=180,in=45] (5.2,-0.8);
% convergence zone label
\node [draw] at (-1.6,-1.6) {\textcolor{NavyBlue}{convergence zone}};
% orientation controller
\node [below] at (9.8,-2) {\textcolor{gray}{orientation}};
\node [below] at (10,-2.6) {\textcolor{gray}{controller}};
% nodes
\draw [fill=NavyBlue] (0,0) circle [radius=0.25];
\draw [fill=ForestGreen] (3,-2) circle [radius=0.25];
\draw [fill=black] (4,-3) circle [radius=0.25];
\draw [fill=red] (6,-2) circle [radius=0.25];
\draw [fill=gray] (10.6,-4.1) circle [radius=0.25];
% nodes name
\node [NavyBlue, below] at (-1,1.6) {$\hat{S}^{(t - 1)}_{i} \rightarrow S^{(t - 1)}_{i}$};
\node [ForestGreen, above] at (3,-1.8) {$S^{(t)}_{i}$};
\node [black, left] at (4,-3.2) {$\hat{S}^{(t)}_{i}$};
\node [red, above right] at (6.4,-2) {$S^{(t + 1)}_{i}$};
\node [gray, above] at (11.5,-4.8) {$X^{(t)}_{i}$};
\end{tikzpicture}
\caption{Overview of Orientation Control.}
\label{fig:7:framework}
\end{figure}

The main idea of our approach is that we inject bounded DP noise into each iteration of the clustering process by applying ExpDP in a controlled orientation of centroids updating, which differs from the existing work where the $\hat{S}^{(t)}_{i}$ was arbitrarily produced by a DP mechanism. Figure~\ref{fig:7:framework} illustrates the overview of our approach. In general, we have three steps to update a set of differentially private centroids at each iteration $t$.
\begin{enumerate}
  \setlength\itemsep{0em}
  \item Run Lloyd's algorithm with past iteration $t-1$ centroid $S^{(t-1)}_{i}$ for a real centroid $S^{(t)}_{i}$ for each cluster $i$. Note that this $S^{(t-1)}_{i}$ is the differentially private centroids $\hat{S}^{(t-1)}_{i}$.
  \item Generate a \textit{sampling zone} by \textit{orientation controller} $X^{(t)}_{i}$ for each cluster $i$;
  \item Sample a differentially private centroid $\hat{S}^{(t)}_{i}$ in this \textit{sampling zone} with ExpDP;
\end{enumerate}

% Differing from the existing work where the $\hat{S}^{(t)}_{i}$ was arbitrarily produced by a DP mechanism, we bound a \textit{sampling zone} to sample the $\hat{S}^{(t)}_{i}$ for the sake of convergence under a DP mechanism. To have a good convergence rate, in our approach, we further define an \textit{orientation controller} (Definition~\ref{def:7:oc}) to control the orientation when sampling the $\hat{S}^{(t)}_{i}$ from the \textit{sampling zone}.

We define a \textit{convergent zone} (for convergence guarantee) and its corresponding \textit{sampling zone} for centroids updating formally in Definition~\ref{def:7:cz}. The specific requirement for the \textit{convergent zone} comes from Lemma~\ref{lem:7:convergence} in next section.
\begin{definition}[Convergent \& Sampling Zones]
\label{def:7:cz}
In $C^{(t)}_{i}$, a \textit{convergent zone} is a set of nodes that Converge Zone = $\{Node\ S:\ ||S - S^{(t)}_{i}|| < ||S^{(t-1)}_{i} - S^{(t)}_{i}||\}$, where $S^{(t)}_{i}$ is the mean of $C^{(t)}_{i}$. A \textit{sampling zone} is a subset of the \textit{convergent zone}.
\end{definition}

\begin{definition}[Orientation Controller]
\label{def:7:oc}
In $C^{(t)}_{i}$, an \textit{orientation controller} is node $X^{(t)}_{i}$ that the differentially private centroids $\hat{S}^{(t)}_{i}$ is randomly sampled by ExpDP according to the orientation $S^{(t)}_{i} \leftarrow X^{(t)}_{i}$.
\end{definition}

The challenge in our scheme to fill the gap is designing a suitable \textit{sampling zone} and an \textit{orientation controller} in the interactive setting to guarantee the convergence and achieve better clustering quality while meeting the same DP requirement as existing work. In the following sections, we propose two types of  \textit{sampling zone} (according to whether we have the knowledge of future centroids movement~\cite{LuZ2019} or not) for our differentially privacy clustering algorithm under this approach to resolve the research challenge.

% pre analysis
\subsection{Preliminary Analysis on Convergence Property}
\label{sec:7:preanalysis}
In this section, we provide the preliminary analysis which helps us build up and analyse our algorithms in the next section. In general, the following properties of the proposed algorithms under our approach would be considered:
\begin{itemize}
  \setlength\itemsep{0em}
  \item The convergence of the proposed algorithms;
  \item The rate of convergence of the proposed algorithms compared with Lloyd's algorithm;
  \item The trade-off between utility and privacy of the proposed algorithms.
\end{itemize}

According to the non-convergence of the existing differentially private $k$-means clustering algorithm, we first study the convergence for a randomised iterative clustering algorithm in Lemma~\ref{lem:7:convergence}.

\begin{lemma}
\label{lem:7:convergence}
A randomised iterative clustering algorithm is convergent if, in $C^{(t)}_{i}$, the sampled $\hat{S}^{(t)}_{i}$ satisfies $||\hat{S}^{(t)}_{i} - S^{(t)}_{i}|| < ||\hat{S}^{(t)}_{i} - S^{(t-1)}_{i}||$ in Euclidean distance, $\forall$ $t$, $i$.
\end{lemma}

\begin{IEEEproof}
In Lloyd's algorithm, after the \textit{re-assignment} step, prior to the \textit{re-centroid} step, we build $C^{(t)}_{i}$ and have $J^{(S_{i}^{(t-1)})} = \sum_{x \in C^{(t)}_{i}}||x - S_{i}^{(t-1)}||^2$, where $S^{(t - 1)}_{i}$ is the mean of $C^{(t-1)}_{i}$ which is used in the \textit{re-assignment} step to generate $C^{(t)}_{i}$. Similarly, after \textit{re-centroid} step, where members in $C^{(t)}_{i}$ did not change, we have $J^{(S_{i}^{(t)})} = \sum_{x \in C^{(t)}_{i}}||x - S_{i}^{(t)}||^2$.

Assuming Euclidean distance between $S_{i}^{(t-1)}$ and $S_{i}^{(t)}$ is $a^{(t)}_{i} = ||S_{i}^{(t-1)} - S_{i}^{(t)}||$, we have 
\begin{equation}
\label{eq:7:gap}
J^{(S_{i}^{(t-1)})} - J^{(S_{i}^{(t)})} = ||C^{(t)}_{i}|| \times (a^{(t)}_{i})^{2}, 
\end{equation}
where $||C^{(t)}_{i}||$ is the number of items in $C^{(t)}_{i}$ (See Section~\ref{sec:7:apdxa} for details of this equation). Note that, in Lloyd's algorithm, $J^{(S_{i}^{(t)})}$ is the minimum cost in $C^{(t)}_{i}$. If we pick a random node $\hat{S}^{(t)}_{i}$ from $C^{(t)}_{i}$ as the centroid for $C^{(t)}_{i}$ which satisfies $||\hat{S}^{(t)}_{i} - S_{i}^{(t)}|| = \delta^{(t)}_{i}a^{(t)}_{i} < ||S_{i}^{(t-1)} - S_{i}^{(t)}|| = a^{(t)}_{i}\ (0 < \delta^{(t)}_{i} < 1)$, then we have $J^{(S_{i}^{(t-1)})} - J^{(\hat{S}^{(t)}_{i})} = ||C^{(t)}_{i}|| \times (1 - (\delta^{(t)}_{i})^{2}) \times (a^{(t)}_{i})^{2} > 0$.

So by updating the centroids to this set $\hat{S}^{(t)} = \{\hat{S}^{(t)}_{1}, \hat{S}^{(t)}_{2}, \dots, \hat{S}^{(t)}_{k}\}$ (rather than the mean of clusters, $S^{(t)}$), the value of every item $\sum_{x \in C_{i}}||x - S_{i}||^2$ can be further decreased, which results in the decrease  of the cost function (Equation~\ref{eq:7:kmeans}). 

In addition, since we have a finite set of all possible clustering solutions (at most $k^{N}$), and we decrease the cost in each iteration of a randomised iterative algorithm, the algorithm satisfies the properties from the above proof must converge (not approach) to a fixed value of the cost function.
% \qed
\end{IEEEproof}

% Based on Lemma~\ref{lem:7:convergence}, we define a \textit{convergent zone} and its corresponding \textit{sampling zone} for centroids updating formally in Definition~\ref{def:7:cz}.
% \begin{definition}[Convergent \& Sampling Zones]
% \label{def:7:cz}
% In $C^{(t)}_{i}$, a \textit{convergent zone} is a set of nodes that Converge Zone = $\{Node\ S:\ ||S - S^{(t)}_{i}|| < ||S^{(t-1)}_{i} - S^{(t)}_{i}||\}$, where $S^{(t)}_{i}$ is the mean of $C^{(t)}_{i}$. A \textit{sampling zone} is a subset of the \textit{convergent zone}.
% \end{definition}

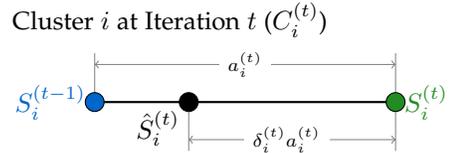
\begin{figure}[!th]
\centering
\begin{tikzpicture}[xscale=0.5, yscale=0.5]
\node [above] at (2,1.5) {Cluster $i$ at Iteration $t$ ($C^{(t)}_{i}$)};
\draw [help lines] (0,0) -- (0,1.3);
\draw [help lines] (2.5,0) -- (2.5,-1.3);
\draw [help lines] (8,1.3) -- (8,-1.3);
\draw [<->, help lines] (0,1) -- (8,1);
\node [fill=white] at (4,1) {\scriptsize{$a^{(t)}_{i}$}};
\draw [<->, help lines] (2.5,-1) -- (8,-1);
\node [fill=white] at (5.1,-1) {\scriptsize{$\delta^{(t)}_{i}a^{(t)}_{i}$}};
\draw [->, thick] (0,0) -- (8,0);
\draw [fill=NavyBlue] (0,0) circle [radius=0.25];
\node [left, NavyBlue] at (0,0) {$S^{(t-1)}_{i}$};
\draw [fill=black] (2.5,0) circle [radius=0.25];
\node [below left, black] at (2.5,0) {$\hat{S}^{(t)}_{i}$};
\draw [fill=ForestGreen] (8,0) circle [radius=0.25];
\node [right, ForestGreen] at (8,0) {$S^{(t)}_{i}$};
\end{tikzpicture}
\caption{Help Figure for Lemma~\ref{lem:7:sameconv} and~\ref{lem:7:iteration}.}
\label{fig:7:dpkm}
\end{figure}

Next, we shall study the convergence and the convergence rate for a special case of $\hat{S}^{(t)}_{i}$ in Lemma~\ref{lem:7:sameconv} and Lemma~\ref{lem:7:iteration}, respectively. This special $\hat{S}^{(t)}_{i}$ (depicts in Figure~\ref{fig:7:dpkm}) is in the line segment of $\overline{S^{(t - 1)}S^{(t)}}$, where $||\hat{S}^{(t)}_{i} - S^{(t)}_{i}|| = \delta^{(t)}_{i} \times ||S^{(t-1)}_{i} - S^{(t)}_{i}||$, $\delta^{(t)}_{i} < 1$. Lemma~\ref{lem:7:sameconv} and Lemma~\ref{lem:7:iteration} assist us to prove the properties of our proposed algorithms in the following sections.

\begin{lemma}
\label{lem:7:sameconv}
Given an algorithm ALG, if we randomly select an $\hat{S}^{(t)}_{i}$ in the line segment of $\overline{S^{(t - 1)}S^{(t)}}$ in $C^{(t)}_{i}$, the convergent degree of ALG is one.
\end{lemma}

\begin{IEEEproof}
We know that the $k$-means clustering problem has a set of local optimal solutions, $\mathbb{S} = \{\mathbf{S}_{1}, \mathbf{S}_{2}, \cdots, \mathbf{S}_{n}\}$, where $\mathbf{S}_{i}$ is one local optimum (the one that Lloyd's algorithm converges to) contains $k$ centroids of the clusters, $\mathbf{S}_{i} = \{S_{i, 1}, S_{i, 2}, \cdots, S_{i, k}\}$. According to Lemma~\ref{lem:7:convergence}, assume ALG is convergent to $\mathbf{\hat{S}} = \{\hat{S}_1, \hat{S}_2, \cdots, \hat{S}_k\} \notin \mathbb{S}$. Then we must have room to further reduce the cost by either the \textit{re-assignment} or the \textit{re-centroid}. Therefore, $\mathbf{\hat{S}}$ is not the set of centroids which makes ALG convergent, unless $\mathbf{\hat{S}} \in \mathbb{S}$. So ALG is convergent to, at least, one local optimum of the $k$-means clustering problem.

We say a set of $k$ nodes (each cluster contributes one node) belongs to a local optimum, $\mathbf{S}_{i}$, if Lloyd's algorithm converges to $\mathbf{S}_{i}$ by taking such a set of nodes as the initial set of centroids. Because the two ends of the line segment $\overline{\mathbf{S}^{(t - 1)}\mathbf{S}^{(t)}}$ belong to the same local optimum, then it is guaranteed that $\mathbf{S}^{(t-1)}$, $\mathbf{S}^{(t)}$, and $\mathbf{\hat{S}}^{(t)}$ always belong to the same local optimum, for all iterations. Therefore, this lemma holds.
% \qed
\end{IEEEproof}

\begin{lemma}
\label{lem:7:iteration}
The algorithm ALG in Lemma~\ref{lem:7:sameconv} has at most $\frac{1}{1 - \delta^{2}}$ times of the iterations of Lloyd's algorithm in the expected case, where $\delta$ is the expectation of $\delta^{(t)}_{i}$, $\delta \in (0, 1)$.
\end{lemma}

\begin{IEEEproof}
Based on Lemma~\ref{lem:7:sameconv}, the overall value difference of Equation~\ref{eq:7:kmeans} from the first iteration to the last iteration, $J = \sum_{i = 1}^{k}J^{S^{(0)}_{i}} - \sum_{i = 1}^{k}J^{S^{(I)}_{i}}$, is the same in both ALG and Lloyd's algorithm, where $I$ is the total iterations. In each iteration, the cost is decreased by two steps: \textit{re-assignment} and \textit{re-centroid}. Then, without loss of generality, we have $J = \sum_{t = 1}^{I}(\Delta^{(ra)}_{t} + \Delta^{(rc)}_{t})$ for Lloyd's algorithm, and $J = \sum_{t = 1}^{\hat{I}}(\hat{\Delta}^{(ra)}_{t} + \hat{\Delta}^{(rc)}_{t})$ for ALG. Because of the properties of Lloyd's algorithm, we know that $\Delta_{t} = \sum_{i = 1}^{k} \Delta_{i}^{(t)}$ for all clusters at iteration $t$. According to Lemma~\ref{lem:7:convergence}, when \textit{re-assignment}, we have $\hat{\Delta}^{(t)}_{i} = (1 - \delta^{2}) \times \Delta^{(t)}_{i}$, where $\delta = \mathbb{E}(\delta^{(t)}_{i})$. So $\hat{\Delta}^{(ra)}_{t} = \sum_{i = 1}^{k} [(1 - \delta^{2}) \times \Delta^{(t)}_{i}] \in [\min_{i = 1}^{k}\{1 - (\delta^{(t)}_{i})^{2}\}, \max_{i = 1}^{k}\{1 - (\delta^{(t)}_{i})^{2}\}] \times \Delta^{(ra)}_{t}$. In the expected case, $\hat{\Delta}^{(ra)}_{t} = (1 - \delta^{2}) \times \Delta^{(ra)}_{t}$, $\delta = \mathbb{E}(\delta^{(t)}_{i})$. In the worst case, $\hat{I} < \frac{1}{\min_{i,t}\{1 - (\delta^{(t)}_{i})^{2}\}} \times I$. As $\hat{\Delta}^{(rc)}_{t} > \Delta^{(rc)}_{i}$, we have 
\begin{equation*}
\begin{split}
J = & (\overline{\Delta^{(ra)}} + \overline{\Delta^{(rc)}}) \times I \\
= & (\overline{\hat{\Delta}^{(ra)}} + \overline{\hat{\Delta}^{(rc)}}) \times \hat{I} \\
> & [(1 - \delta^{2})\overline{\Delta^{(ra)}} + \overline{\Delta^{(rc)}}] \times \hat{I}\\
> & (1 - \delta^{2}) \times (\overline{\Delta^{(ra)}} + \overline{\Delta^{(rc)}}) \times \hat{I}.
\end{split}
\end{equation*}
Therefore, $\hat{I} < \frac{1}{1 - \delta^{2}} \times I$ in the expected case.
% \qed
\end{IEEEproof}

% sampling zone
\section{\textit{Sampling Zone} Design}
\label{sec:7:samplingzone}
In this section, we first discuss the rules for building a \textit{sampling zone}, then show the two designs of \textit{sampling zone} we propose.

\subsection{Design Rules}
Ideally, in our \textit{convergent zone}, when applying LapDP, the probability of a node $S$ as the $\hat{S}^{(t)}_{i}$ need follow a monotonous decreasing function of the distance between $S$ and $S^{(t)}_{i}$. A truncated LapDP~\cite{ANDRES2013} would be a straightforward way to achieve our goal. That is, once the random noise of LapDP is outside the \textit{convergent zone}, we truncate it to the border of \textit{convergent zone}. However, this truncated LapDP will introduce a contradiction against the above ideal case. Because the nodes in the border of the \textit{convergent zone} may have a higher probability (sum of the probabilities of the nodes outside the \textit{convergent zone}) than the ones closer to the $S^{(t)}_{i}$. Therefore, in this paper, we apply the ExpDP in the \textit{convergent zone} (in fact, in the \textit{sampling zone}) to sample the $\hat{S}^{(t)}_{i}$.

When designing a \textit{sampling zone} under our approach, we should follow the following rules. Firstly, there should be a single \textit{sampling zone} in $C^{(t)}_{i}$ for all parties: the trusted data curator and the adversaries. Otherwise the differences among the \textit{sampling zone}s in different parties will result in significant differences among their clustering results, which could be used for privacy inference. Secondly, the single \textit{sampling zone} should not have an explicit relationship to  $S^{(t)}_{i}$, the real mean of $C^{(t)}_{i}$, since otherwise the adversary may easily learn the expected value of $S^{(t)}_{i}$. With high probability, the expectation can be used as the real value. Thirdly, to control the convergence orientation, the \textit{orientation controller} should be involved when building the \textit{sampling zone}. 

Based on the above discussions of the \textit{sampling zone} and our research challenges presented in Section~\ref{sec:7:overview}, we shall apply two strategies for the \textit{orientation controller} to build two types of  \textit{sampling zone}  in the following sections. The major difference between the two strategies is that whether we use the past knowledge only or both past and future knowledge~\cite{LuZ2019} of the cluster centroids as the \textit{orientation controller} for the \textit{sampling zone}. Such a difference results in variant clustering qualities and convergence rate.

% orientation controller
\subsection{Orientation Control with Past Knowledge}
% Noting that the \textit{sampling zone} with the future knowledge of the cluster centroids will involve more computations in each iteration, so to save computational resources, instead of requiring global knowledge, we will use only local knowledge of iteration $i$ to build the \textit{sampling zone} of iteration $i$. We observe that, in $C^{(t)}_{i}$, the orientation of $S^{(t-1)}_{i} \rightarrow S^{(t)}_{i}$ indicates a trend of cluster centroids movement. Therefore, without the assistant of future knowledge of the exact centroids movement orientation (i.e., $S^{(t)}_{i} \rightarrow S^{(t+1)}_{i}$), the \textit{orientation controller} could be the point of intersection of the \textit{convergent zone}'s borderline and the line $S^{(t-1)}_{i}S^{(t)}_{i}$. However, since such a point of intersection has an explicit relationship to $S^{(t-1)}_{i}$ and $S^{(t)}_{i}$, we cannot use it as the \textit{orientation controller} directly. To solve this problem, we simply shift this point of intersection with a random angle to have our \textit{orientation controller}, $X^{(t)}_{i}$. Because we still want the $X^{(t)}_{i}$ is as close as the point of intersection, we use the following probability function for sampling an angle $\gamma$: $\Pr[\gamma^{(t)}_{i} = r] \propto \exp(1 - 2|r|/\pi)$, $r \in [0, \pi/2]$.
We observe that, in $C^{(t)}_{i}$, the past knowledge that the orientation of $S^{(t-1)}_{i} \rightarrow S^{(t)}_{i}$ indicates a trend of cluster centroids movement. Therefore, the \textit{orientation controller} could be the point of intersection of the \textit{convergent zone}'s borderline and the line $S^{(t-1)}_{i}S^{(t)}_{i}$. However, since such a point of intersection has an explicit relationship to $S^{(t-1)}_{i}$ and $S^{(t)}_{i}$, we cannot use it as the \textit{orientation controller} directly. To solve this problem, we simply shift this point of intersection with a random angle to have our \textit{orientation controller}, $X^{(t)}_{i}$. Because we still want the $X^{(t)}_{i}$ is as close as the point of intersection, we use the following probability function for sampling an angle $\gamma$: $\Pr[\gamma^{(t)}_{i} = r] \propto \exp(1 - 2|r|/\pi)$, $r \in [0, \pi/2]$.

\subsection{Orientation Control with Past and Future Knowledge}
Clearly, the \textit{sampling zone} with the past knowledge of the cluster centroids cannot guarantee the convergence orientation towards to the convergence of Lloyd's algorithm over the iterations, which will result a poor convergence degree, so to improve the convergence quality, we use the centroids movement in the future iterations as the orientation for centroids updating. As we know, Lloyd algorithm approaches to a local optimum of the $k$-means clustering problem through iterations. If we use the final/convergent centroid, $S^{(t + r_{t})}_{i}$, as the \textit{orientation controller} in $C^{(t)}_{i}$, we can provide clustering quality in our random mechanism (i.e., the ExpDP) as much as possible. Note that, in $C^{(t)}_{i}$, $S^{(t + r_{t})}_{i}$ is the future knowledge of the cluster centroids. However, taking such an $S^{(t + r_{t})}_{i}$ means we have to further run Lloyd's algorithm for $r_{t}$ iterations in $C^{(t)}_{i}$, which will result in a large rate of convergence when our differentially private algorithm converges. Therefore, considering the computational cost, we choose the \textit{orientation controller}, $X^{(t)}_{i}$, as $S^{(t + 1)}_{i}$.

% sampling zone implementation
\section{Proposed Algorithm and Its Analysis}
\label{sec:7:algorithm}
In this section, we show our proposed differentially private $k$-means clustering algorithm with guaranteed convergence and the analysis on its convergence, convergence rate, and differential privacy.

% main algorithm
\subsection{The Clustering Algorithm}
\begin{figure*}[!th]
\centering
% sampling zone 1
\subfloat[Posterior \textit{Sampling Zone} (Past Knowledge).]{
\begin{tikzpicture}[xscale=0.5, yscale=0.5]
% blue circle
\draw [dashed, NavyBlue] (0,0) circle [radius=6];
% brown circle
\draw [dashed, myBrown] (0.7,-3.5) circle [radius=2.448];
% blue circle label
\node [right, draw, fill=white] at (-7,4.5) {\textcolor{NavyBlue}{convergent zone}};
% link label to circle, blue
\draw [NavyBlue, ->] (-6.5,4) to [out=270,in=180] (-5.5,2.7);
% brown circle label
\node [right, draw, fill=white] at (-4.3,2.5) {\textcolor{myBrown}{sampling zone}};
% link label to circle, brown
\draw [myBrown, ->] (-3.5,1.95) to [out=270,in=180] (-1.8,-3);
% t-1 to t
\draw [->, thick] (4.24,4.24) -- (0.177,0.177);
% t to y
\draw [thick] (0,0) -- (-4.24,-4.24);
% t to x
\draw [thick] (0,0) -- (1.18,-5.9);
% arc for gamma
\draw (-0.25,-0.25) arc [radius=0.25, start angle=200, end angle=290];
\draw (-0.35,-0.35) arc [radius=0.35, start angle=200, end angle=290];
% gamma
\node [below] at (-0.25, -0.35) {\scriptsize{$\gamma^{(t)}_{i}$}};
% help lines for t to p
\draw [help lines] (0,0) -- (3,0.6);
\draw [help lines] (0.7,-3.5) -- (2.2,-3.2);
\draw [help lines, <->] (1.2,6/25) -- (1.9,-163/50);
\node [fill=white] at (4.5,1) {\scriptsize{$\lambda^{(t)}_{i}b^{(t)}_{i}$}};
\draw [help lines, dashed, ->] (4.5,0.4) to [out=270,in=0] (1.4,-0.6);
% help lines for t to x
\draw [help lines] (1.18,-5.9) -- (4.1,-5.315);
\draw [help lines, <->] (2.7,27/50) -- (3.88,-5.36);
\node [fill=white] at (3.2,-1.5) {\scriptsize{$b^{(t)}_{i}$}};
% point t-1
\draw [fill=NavyBlue] (4.24,4.24) circle [radius=0.25];
\node [above right, NavyBlue] at (4.2,4.2) {$S^{(t-1)}_{i}$};
% point t
\draw [fill=ForestGreen] (0,0) circle [radius=0.25];
\node [above, ForestGreen] at (0,0.1) {$S^{(t)}_{i}$};
% point y
\draw [fill=myBrown] (-4.24,-4.24) circle [radius=0.25];
\node [above, myBrown] at (-4.24,-4) {$Y^{(t)}_{i}$};
% point x
\draw [fill=red] (1.18,-5.9) circle [radius=0.25];
\node [above left, red] at (1.18,-5.9) {$X^{(t)}_{i}$};
% point p
\draw [fill=myGrey] (0.7,-3.5) circle [radius=0.15];
\node [below right, myGrey] at (0.7,-3.3) {$P^{(t)}_{i}$};
% point m
\draw [fill=gray] (0.59,-2.95) circle [radius=0.15];
\node [above right, gray] at (0.3,-3) {\scriptsize{$M^{(t)}_{i}$}};
\end{tikzpicture}
\label{fig:7:posteriorsampling}
}
% \hfill
% sampling zone 2
\subfloat[Prior \textit{Sampling Zone} (Past and Future Knowledge).]{
\begin{tikzpicture}[xscale=0.5, yscale=0.5]
% blue circle
\draw [dashed, NavyBlue] (0,0) circle [radius=6];
% brown circle
\draw [dashed, myBrown] (0.6,-3.0) circle [radius=2.04];
% blue circle label
\node [right, draw, fill=white] at (-7,4.5) {\textcolor{NavyBlue}{convergent zone}};
% link label to circle, blue
\draw [NavyBlue, ->] (-6.5,4) to [out=270,in=180] (-5.5,2.7);
% brown circle label
\node [right, draw, fill=white] at (-4.3,2.5) {\textcolor{myBrown}{sampling zone}};
% link label to circle, brown
\draw [myBrown, ->] (-3.5,1.95) to [out=270,in=180] (-1.5,-3);
% t-1 to t
\draw [->, thick] (4.24,4.24) -- (0.177,0.177);
% t to t+1
\draw [->, thick] (0,0) -- (0.951,-4.755);
% help lines for t to p
\draw [help lines] (0,0) -- (3,0.6);
\draw [help lines] (0.6,-3) -- (2.2,-67/25);
\draw [help lines, <->] (1.2,6/25) -- (1.8,-69/25);
\node [fill=white] at (4.5,1) {\scriptsize{$\lambda^{(t)}_{i}b^{(t)}_{i}$}};
\draw [help lines, dashed, ->] (4.5,0.4) to [out=270,in=0] (1.4,-0.6);
% help lines for t to t+1
\draw [help lines] (1,-5) -- (4,-22/5);
\draw [help lines, <->] (2.7,27/50) -- (3.7,-223/50);
\node [fill=white] at (3.2,-1.5) {\scriptsize{$b^{(t)}_{i}$}};
% point t-1
\draw [fill=NavyBlue] (4.24,4.24) circle [radius=0.25];
\node [above right, NavyBlue] at (4.2,4.2) {$S^{(t-1)}_{i}$};
% point t
\draw [fill=ForestGreen] (0,0) circle [radius=0.25];
\node [above, ForestGreen] at (0,0.1) {$S^{(t)}_{i}$};
% point t+1
\draw [fill=red] (1,-5) circle [radius=0.25];
\node [below left, red] at (1,-4.8) {$S^{(t+1)}_{i} \rightarrow X^{(t)}_{i}$};
% point p
\draw [fill=myGrey] (0.6,-3) circle [radius=0.15];
\node [below right, myGrey] at (0.6,-2.8) {$P^{(t)}_{i}$};
% point m
\draw [fill=gray] (0.5,-2.5) circle [radius=0.15];
\node [above right, gray] at (0.25,-2.57) {\scriptsize{$M^{(t)}_{i}$}};
\end{tikzpicture}
\label{fig:7:priorsampling}
}
\caption{The General Idea of the \textit{Sampling Zone}.}
\label{fig:7:samplingzone}
\end{figure*}
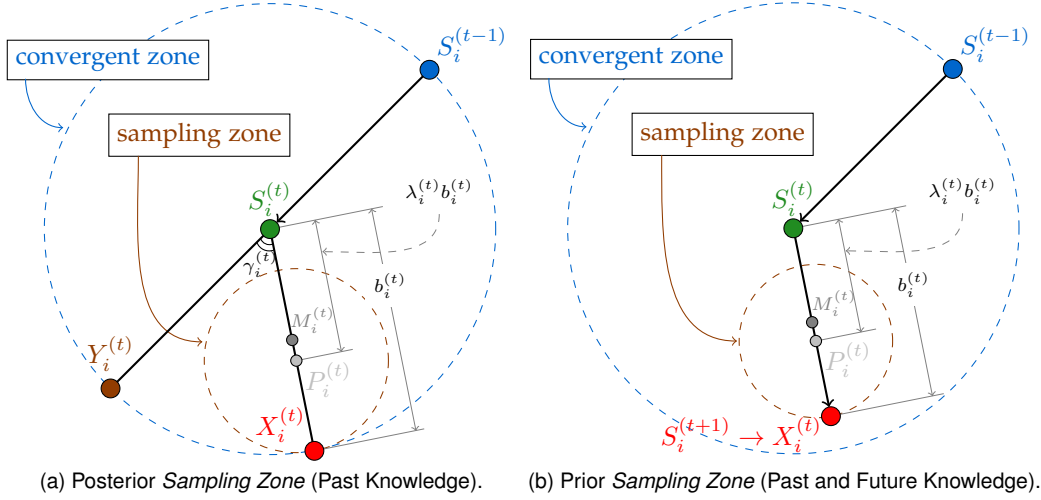

The first step of our algorithm is \textit{sampling zone} generation. We generate our \textit{sampling zone} by computing its centre and radius, respectively. The centre of the \textit{sampling zone}, $P^{(t)}_{i}$, is determined by a random number $\lambda^{(t)}_{i} \in (1/2, 1)$ which is the off-set in $\overline{S^{(t)}_{i}X^{(t)}_{i}}$. Because a larger \textit{sampling zone} provides more choices for the $\hat{S}^{(t)}_{i}$, we use the following probability function for sampling $\lambda^{(t)}_{i}$ as $\Pr[P^{(t)}_{i}] = \Pr[\lambda^{(t)}_{i} = r] \propto \exp(2 - 2r) = p, r \in (1/2, 1)$. The radius of the \textit{sampling zone}, $r^{(t)}_{i} = ||X^{(t)}_{i} - P^{(t)}_{i}||$. In this paper, depending on whether the we use past knowledge only or both past and future knowledge, we name the \textit{sampling zone} as prior \textit{sampling zone} (past knowledge) and posterior \textit{sampling zone} (past and future knowledge). Algorithm~\ref{alg:7:samplingzone} shows how we build the \textit{sampling zone} with either past knowledge or past plus future knowledge of the cluster centroids. Figure~\ref{fig:7:samplingzone} depicts the key idea of our building process of the \textit{sampling zone}.
% sampling zone

Second, once having the \textit{sampling zone}, each party samples their own $\hat{S}^{(t)}_{i}$ from this \textit{sampling zone} with the ExpDP. In the implementation, we sample the $\hat{S}^{(t)}_{i}$ by sampling a pair $(\delta^{(t)}_{i} = \frac{||S^{(t)}_{i} - \hat{S}^{(t)}_{i}||}{||S^{(t)}_{i} - S^{(t+1)}_{i}||}, \alpha^{(t)}_{i} = \angle \hat{S}^{(t)}_{i} S^{(t)}_{i} S^{(t+1)}_{i})$, where $\delta^{(t)}_{i} \in (0, 1)$, $\alpha^{(t)}_{i} \in (-\pi / 2, \pi / 2)$. Because an $\hat{S}^{(t)}_{i}$, that is close to the $S^{(t)}_{i}$, has better clustering quality for iterations in the interactive setting, that is, the scoring function should be monotonous decreasing to both $\delta^{(t)}_{i}$ and $\alpha^{(t)}_{i}$. In this paper, we use the following scoring function for the pair $(\delta^{(t)}_{i}, \alpha^{(t)}_{i})$ because of its simplicity: $q(\delta^{(t)}_{i}, \alpha^{(t)}_{i}) = (1 - \delta^{(t)}_{i}) + (1 - 2|\alpha^{(t)}_{i}|/\pi)$. It is easy to see that the local sensitivity of the scoring function is 2, i.e. $\Delta q = 2$.
\begin{figure}[!th]
\centering
\begin{tikzpicture}[xscale=0.5, yscale=0.5]
%brown circle
\draw [dashed, myBrown] (0.6,-3.0) circle [radius=2.04];
%brown circle label
\node [right, draw, fill=white] at (-4.3,2.5) {\textcolor{myBrown}{sampling zone}};
%link label to circle, brown
\draw [myBrown, ->] (-3.5,1.95) to [out=270,in=175] (-1,-4.4);
%t to x
\draw [thick] (0,0) -- (0.951,-4.755);
%t to hat t
\draw (0,0) -- (-1,-3);
%arc for alpha
\draw (-0.126,-0.379) arc [radius=0.5, start angle=251.5, end angle=278];
\draw (-0.174,-0.523) arc [radius=0.65, start angle=251.5, end angle=278];
%alpha
\node [below] at (-0.05, -0.5) {\scriptsize{$\alpha$}};
%help lines for t to hat t
\draw [help lines] (0,0) -- (-1.3,13/30);
\draw [help lines] (-1,-3) -- (-2.3,-77/30);
\draw [help lines, <->] (-1,1/3) -- (-2,-8/3);
\node [fill=white] at (-2,-1) {\scriptsize{$\delta^{(t)}_{i}b^{(t)}_{i}$}};
%help lines for t to p
\draw [help lines] (0,0) -- (3,0.6);
\draw [help lines] (0.6,-3) -- (2.2,-67/25);
\draw [help lines, <->] (1.2,6/25) -- (1.8,-69/25);
\node [fill=white] at (4.5,1) {\scriptsize{$\lambda^{(t)}_{i}b^{(t)}_{i}$}};
\draw [help lines, dashed, ->] (4.5,0.4) to [out=270,in=0] (1.4,-0.6);
%help lines for t to x
\draw [help lines] (1,-5) -- (4,-22/5);
\draw [help lines, <->] (2.7,27/50) -- (3.7,-223/50);
\node [fill=white] at (3.5,-1.5) {\scriptsize{$b^{(t)}_{i}$}};
%point t
\draw [fill=ForestGreen] (0,0) circle [radius=0.25];
\node [above, ForestGreen] at (0,0.1) {$S^{(t)}_{i}$};
%point x
\draw [fill=red] (1,-5) circle [radius=0.25];
\node [below left, red] at (1,-4.8) {$X^{(t)}_{i}$};
%point p
\draw [fill=myGrey] (0.6,-3) circle [radius=0.15];
\node [below right, myGrey] at (0.6,-2.8) {$P^{(t)}_{i}$};
%point m
\draw [fill=gray] (0.5,-2.5) circle [radius=0.15];
\node [above right, gray] at (0.15,-2.57) {\scriptsize{$M^{(t)}_{i}$}};
%point hat
\draw [fill=black] (-1,-3) circle [radius=0.25];
\node [right] at (-1.3,-3.8) {$\hat{S}^{(t)}_{i}$};
\end{tikzpicture}
\caption{The Key Idea of Algorithm~\ref{alg:7:edpkm}: Centroids Updating.}
\label{fig:7:edpkm}
\end{figure}
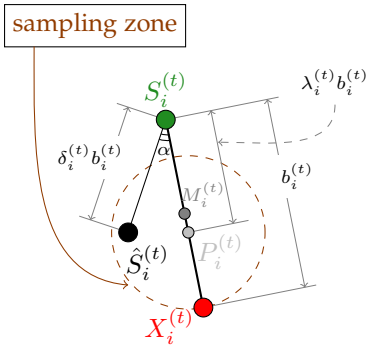

Finally, when the clusters converge (to a real local optimum as Lloyd's algorithm), we apply the LapDP to inject noise to the final clustering result. Specifically, to have good clustering quality, we inject the Laplace noise to the counts when calculating the mean of each cluster (Line 12 in Algorithm~\ref{alg:7:edpkm}). The local sensitivity of this counting function is 1. Algorithm~\ref{alg:7:edpkm} shows how our approach works.

\begin{algorithm}[!th]
% \scriptsize
\caption{Differentially Private $k$-Means Clustering.}
\label{alg:7:edpkm}
\SetKwInOut{Input}{Input}
\SetKwInOut{Output}{Output}

\Input{
$X = \{x_{1}, x_{2}, \dots, x_{N}\}$: dataset in size $N$.\\
$k$: number of clusters ($< N$).\\
$\epsilon^{(t)}_{i}$: privacy budget for Cluster $i$ at Iteration $t$, $C^{(t)}_{i}$.\\
$\epsilon_{0}$: privacy budget for the final output.\\
$\Pr[P^{(t)}_{i}]$: probability to generate $SamplingZone^{(t)}_{i}$ for $C^{(t)}_{i}$.\\
$q$: scoring function for the ExpDP when sampling the $\hat{S}^{(t)}_{i}$.
}
\Output{$\mathbf{S}$: set of the final $k$ centroids.}
\BlankLine
Initialisation: Uniformly sample $k$ initial centroids $\mathbf{S}^{(0)} = (S^{(0)}_{1}, S^{(0)}_{2}, \dots, S^{(0)}_{k})$ from $X$\;
\While{clusters do not converge}{
  \For{each Cluster $i$ at Iteration $t$}{
    $C^{(t)}_{i}$ $\leftarrow$ assign each $x_j$ to its closest centroid $S^{(t-1)}_{i}$\;
    $S^{(t)}_{i}$ $\leftarrow$ mean of $C^{(t)}_{i}$\;
    $SamplingZone^{(t)}_{i}$ $\leftarrow$ run Algorithm~\ref{alg:7:samplingzone}\;
    $\hat{S}^{(t)}_{i}$ $\leftarrow$ sample from $SamplingZone^{(t)}_{i}$ by the ExpDP with $q$ and $\epsilon^{(t)}_{i}$\;
    $S^{(t)}_{i}$ $\leftarrow$ $\hat{S}^{(t)}_{i}$\;
    Publish: $SamplingZone^{(t)}_{i}$, $q$, $\epsilon^{(t)}_{i}$, $S^{(t)}_{i}$(optional)\;
  }
}
$\mathbf{S}$ $\leftarrow$ add noise to $\mathbf{S}^{(t)}$ by the LapDP with $\epsilon_{0}$, publish $\epsilon_{0}$\;
\end{algorithm}

% Alg for sampling zone
\begin{algorithm}[!th]
% \scriptsize
\caption{The \textit{Sampling Zone} Generator.}
\label{alg:7:samplingzone}
\SetKwInOut{Input}{Input}
\SetKwInOut{Output}{Output}

\Input{
$S^{(t)}_{i}$: mean of $C^{(t)}_{i}$.\\
$S^{(t-1)}_{i}$: mean of $C^{(t-1)}_{i}$.\\
$\Pr[P^{(t)}_{i}]$: probability to generate $SamplingZone^{(t)}_{i}$ for $C^{(t)}_{i}$.\\
$\Pr[\gamma^{(t)}_{i}]$: probability to generate angle $\gamma^{(t)}_{i}$.\\
$useFuture$: future knowledge of cluster centroids.
}
\Output{$SamplingZone^{(t)}_{i}$.}
\eIf{useFuture is yes}{
  $X^{(t)}_{i}$ $\leftarrow$ mean of $C^{(t+1)}_{i}$ based on $C^{(t)}_{i}$\;
}
{
  $Y^{(t)}_{i}$ $\leftarrow$ the point of intersection of the \textit{convergent zone}'s borderline and the line $S^{(t-1)}_{i}S^{(t)}_{i}$\;
  $\gamma^{(t)}_{i}$ $\leftarrow$ sample by $\Pr[\gamma^{(t)}_{i}]$\;
  $X^{(t)}_{i}$ $\leftarrow$ shift $Y^{(t)}_{i}$ on \textit{convergent zone}'s borderline with angle $\gamma^{(t)}_{i}$\;
}
$M^{(t)}_{i}$ $\leftarrow$ midpoint of the line segment $\overline{S^{(t)}_{i}X^{(t)}_{i}}$\;
$P^{(t)}_{i}$ $\leftarrow$ sample from the line segment $\overline{M^{(t)}_{i}X^{(t)}_{i}}$ by $\Pr[P^{(t)}_{i}]$\;
$SamplingZone^{(t)}_{i}$ $\leftarrow$ centre: $P^{(t)}_{i}$, radius: $r^{(t)}_{i} = ||X^{(t)}_{i} - P^{(t)}_{i}||$\;
\end{algorithm}

% analysis
\subsection{Proof of Convergence and Differential Privacy}
According to Lemma~\ref{lem:7:convergence}, \ref{lem:7:sameconv}, and \ref{lem:7:iteration}, we have Theorem~\ref{thm:7:sameconv2}, \ref{thm:7:iteration2}, and \ref{thm:7:sameconv}, \ref{thm:7:iteration} to study the convergence and the convergence rate of Algorithm~\ref{alg:7:edpkm}, respectively. Theorem~\ref{thm:7:privacy} studies the privacy bound of Algorithm~\ref{alg:7:edpkm}.

\begin{theorem}
\label{thm:7:sameconv2}
Algorithm~\ref{alg:7:edpkm} (\textit{sampling zone} with past knowledge) has convergence degree at least $1/m$ where $m$ is the number of local optima of Lloyd's  algorithm for a given dataset.
\end{theorem}
\begin{IEEEproof}
The convergent orientation is not determined when the \textit{sampling zone} relies on the past knowledge. With uniform distribution for the orientation, if there are $m$ local optimum of $k$-means problem for a given dataset, the convergent degree will be at least $1/m$ in this case.
\end{IEEEproof}

\begin{theorem}
\label{thm:7:iteration2}
Algorithm~\ref{alg:7:edpkm} (\textit{sampling zone} with past knowledge) converges in at most two times of the iterations of Lloyd's algorithm in the expected case.
%, if Algorithm~\ref{alg:7:edpkm} converges to the same centroids as Lloyd's algorithm.
\end{theorem}
\begin{IEEEproof}
According to Lemma~\ref{lem:7:convergence} and Theorem~\ref{thm:7:iteration}, the key points for analysing the convergence rate are the length of $||S^{(t)}_{i} - S^{(t+1)}_{i}||$ and $||\hat{S}^{(t)}_{i} - S^{(t+1)}_{i}||$. That is,
$$\hat{I} < \frac{||S^{(t)}_{i} - S^{(t+1)}_{i}||^{2}}{||\hat{S}^{(t)}_{i} - S^{(t+1)}_{i}||^{2}} \times I.$$

Because, in this \textit{sampling zone} (with past knowledge), we cannot determine the angle $\alpha$ in Figure~\ref{fig:7:proofiter} to figure out the explicit expression for $||S^{(t)}_{i} - S^{(t+1)}_{i}||$ and $||\hat{S}^{(t)}_{i} - S^{(t+1)}_{i}||$, we simply use the triangle inequality to find the upper bound of $\hat{I}$. According to the triangle inequality, $(1 - \delta^{(t)}_{i})^{2} < \frac{||\hat{S}^{(t)}_{i} - S^{(t+1)}_{i}||^{2}}{||S^{(t)}_{i} - S^{(t+1)}_{i}||^{2}} < (1 + \delta^{(t)}_{i})^{2}$. Note that, in this case, $\delta^{(t)}_{i}$ may be greater than 1. So we have,
$$\hat{I} < \frac{1}{(1 - \delta^{(t)}_{i})^{2}} \times I.$$

Because our \textit{sampling zone} is a subset of the \textit{convergent zone}, we must have $\hat{I} \leq 2I$. Then we have $\hat{I} \leq \min\{2, \frac{1}{(1 - \delta^{(t)}_{i})^{2}}\}$. Note that, when $\delta^{(t)}_{i} \leq \frac{2 - \sqrt{2}}{2}$ or $\delta^{(t)}_{i} \geq \frac{2 + \sqrt{2}}{2}$, $\frac{1}{(1 - \delta^{(t)}_{i})^{2}} \leq 2$.
% \qed
\end{IEEEproof}

\begin{theorem}
\label{thm:7:sameconv}
Given a set of initial centroids, Algorithm~\ref{alg:7:edpkm} (\textit{sampling zone} with past and future knowledge) has convergence degree 1, i.e., converges to the same (final) centroids as Lloyd's algorithm, with at least $1-\frac{1}{2}(\frac{m}{n})^{\frac{d-1}{d}}$ probability, where $n$ is the number of items in a dataset $D$, $d$ is the dimension of an item, $m$ is the number of local optima of Lloyd's algorithm on dataset $D$.
\end{theorem}

\begin{IEEEproof}
In Algorithm~\ref{alg:7:edpkm}, because each \textit{sampling zone} is a subset of a \textit{convergent zone}, according to Lemma~\ref{lem:7:convergence}, Algorithm~\ref{alg:7:edpkm} is convergent. According to Lemma~\ref{lem:7:sameconv}, any arbitrary set of $k$ nodes as the initial set of centroids must converge to a local optimum in Lloyd's algorithm. However, for some sets of $k$ nodes as the initial centroids, they may belong to different local optimum. Such nodes appear at the border area between two local optimums. Assume a dataset $D$ contains $n$ items, each item has $d$ dimensions, the average distance between two items is $l$, then the overall size of the space of $D$ is $nl^{d}$. The overall size of the border area space is $\frac{m}{2} \times (\frac{nl^{d}}{m})^{\frac{1}{d}} \times [2(b-\lambda b)]^{(d-1)}$, where $m$ is the number of local optimum. Then we have at least $1 - \frac{\frac{m}{2} \times (\frac{nl^{d}}{m})^{\frac{1}{d}} \times [2(b-\lambda b)]^{(d-1)}}{nl^{d}} \geq 1-\frac{1}{2}(\frac{m}{n})^{\frac{d-1}{d}}$ probability to not sample the initial nodes from the border area. Since $||S^{(t+1)}_{i} - S^{(t)}_{i}|| < ||S^{(t-1)}_{i} - S^{(t)}_{i}||$, when $t > 1$, all the sets of $k$ nodes from $SamplingZone^{(1)}_{i}$ belong to same local optimum. Therefore, based on Lemma~\ref{lem:7:sameconv}, this theorem holds.
% \qed
\end{IEEEproof}

\begin{theorem}
\label{thm:7:iteration}
Algorithm~\ref{alg:7:edpkm} (\textit{sampling zone} with past and future knowledge) converges in at most $\frac{2}{-\delta^{2} + 2\delta \cos \alpha + 1} \in (1, 2)$ times of the iterations of Lloyd's algorithm in the expected case, where $\delta$ and $\alpha$ are the expectations of $\delta^{(t)}_{i}$ and $\alpha^{(t)}_{i}$.
\end{theorem}

\begin{IEEEproof}
According to Lemma~\ref{lem:7:iteration}, the total number of iteration of Algorithm~\ref{alg:7:edpkm} depends on the distance $||\hat{S}^{(t)}_{i} - S^{(t+1)}_{i}||$. By building a help figure (Figure~\ref{fig:7:proofiter}), we have:
\begin{equation*}
\begin{split}
 & ||\hat{S}^{(t)}_{i} - S^{(t+1)}_{i}||^{2} \\
= & ||\hat{S}^{(t)}_{i} - Temp||^{2} + ||Temp - S^{(t+1)}_{i}||^{2} \\
= & ||\hat{S}^{(t)}_{i} - Temp||^{2} + (||S^{(t)}_{i} - S^{(t+1)}_{i}|| \\
& - ||Temp - S^{(t)}_{i}||)^{2} \\
% = & (\delta^{(t)}_{i}b^{(t)}_{i} \sin \alpha^{(t)}_{i})^{2} + (b^{(t)}_{i} - \delta^{(t)}_{i}b^{(t)}_{i} \cos \alpha^{(t)}_{i})^{2}\\
= & [(\delta^{(t)}_{i})^{2} - 2\delta^{(t)}_{i} \cos \alpha^{(t)}_{i} + 1](b^{(t)}_{i})^{2}.
\end{split}
\end{equation*}

\begin{figure}[!th]
\centering
\begin{tikzpicture}[xscale=0.6, yscale=0.6]
\draw [help lines] (0,0) -- (2,-4);%t to t+1
\draw [help lines] (0,0) -- (-1,-3);%t to hat t
\draw [help lines] (-1,-3) -- (2,-4);%hat t to t+1
\draw [help lines, dashed] (-1,-3) -- (1,-2);
\draw (1.2,-2.4) -- (0.8,-2.6) -- (0.6,-2.2);
\draw (-0.126,-0.379) arc [radius=0.4, start angle=251.5, end angle=296.5];%arc for alpha
\draw (-0.174,-0.523) arc [radius=0.55, start angle=251.5, end angle=296.5];%arc for alpha
\node [below] at (0.1, -0.5) {$\alpha$};
\draw [help lines] (0,0) -- (-0.7,7/30);
\draw [help lines] (-1,-3) -- (-1.7,-83/30);
\draw [help lines, <->] (-0.5,1/6) -- (-1.5,-17/6);
\node [fill=white] at (-1.7,-1) {\scriptsize{$\delta^{(t)}_{i}b^{(t)}_{i}$}};
\draw [help lines] (0,0) -- (1.8,0.9);
\draw [help lines] (2,-4) -- (3.8,-3.1);
\draw [help lines, <->] (1.4,0.7) -- (3.4,-3.3);
\node [fill=white, right] at (2,-1.4) {\scriptsize{$b^{(t)}_{i}$}};
\draw [fill=ForestGreen] (0,0) circle [radius=0.25];
\node [above, ForestGreen] at (0,0.1) {$S^{(t)}_{i}$};
\draw [fill=red] (2,-4) circle [radius=0.25];
\node [below right, red] at (2,-3.5) {$S^{(t+1)}_{i}$};
\draw [fill=black] (-1,-3) circle [radius=0.25];
\node [below left] at (-1,-3) {$\hat{S}^{(t)}_{i}$};
\draw [fill=gray] (1,-2) circle [radius=0.2];
\node [right, gray] at (1,-2) {\scriptsize{$Temp$}};
\end{tikzpicture}
\caption{Help Figure for Proof of Theorem~\ref{thm:7:iteration}.}
\label{fig:7:proofiter}
\end{figure}
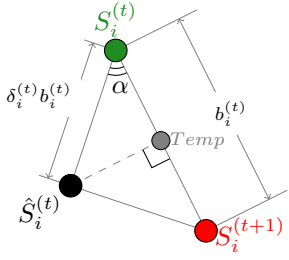

Then we have the ratio
\begin{equation*}
\frac{||\hat{S}^{(t)}_{i} - S^{(t+1)}_{i}||}{||S^{(t)}_{i} - S^{(t+1)}_{i}||} = \sqrt{(\delta^{(t)}_{i})^{2} - 2\delta^{(t)}_{i} \cos \alpha^{(t)}_{i} + 1}.
\end{equation*}

In Algorithm~\ref{alg:7:edpkm}, we calculate the centroid $S^{(t+1)}_{i}$ at iteration $t$, so it is supposed to have a $\Delta_{t} + \Delta_{t+1}$ change for the cost value. However, by applying the similar idea from Lemma~\ref{lem:7:iteration}, what we have is
\begin{equation*}
\begin{split}
& \hat{\Delta}_{t} + \hat{\Delta}_{t+1} \\
> & \Delta_{t} + (1 - (\sqrt{\delta^{2} - 2\delta \cos \alpha + 1})^{2}) \times \Delta_{t+1}\\
= & \Delta_{t} + (-\delta^{2} + 2\delta \cos \alpha) \times \Delta_{t+1},
\end{split}
\end{equation*}
where $\hat{\Delta}_{t} = \Delta_{t}$.

Recall how Algorithm~\ref{alg:7:edpkm} converges, half iterations decrease the cost function as $\hat{\Delta_{t}}$, half iterations do so as $\Delta_{t+1}$. So assume $\hat{I} = T \times I$ the overall decreasing of the cost function is
\begin{equation*}
\begin{split}
& \frac{1}{2}\hat{I}\overline{\hat{\Delta}_{t}} + \frac{1}{2}\hat{I}\overline{\hat{\Delta}_{t+1}} \\
> & \frac{1}{2}\hat{I}\overline{\Delta_{t}} + \frac{1}{2}(-\delta^{2} + 2\delta \cos \alpha) \times \hat{I}\overline{\Delta_{t+1}}\\
= & \frac{1}{2}T \times I\overline{\Delta_{t}} + \frac{1}{2}(-\delta^{2} + 2\delta \cos \alpha)T \times I\overline{\Delta_{t+1}}.
\end{split}
\end{equation*}
Then we have $\frac{1}{2}T + \frac{1}{2}(-\delta^{2} + 2\delta \cos \alpha)T < 1$, so $T < \frac{2}{-\delta^{2} + 2\delta \cos \alpha + 1}$. Note that since $||\hat{S}^{(t)}_{i} - S^{(t+1)}_{i}|| < ||S^{(t)}_{i} - S^{(t+1)}_{i}||$, $\forall$ $t$ and $i$, $\sqrt{(\delta^{(t)}_{i})^{2} - 2\delta^{(t)}_{i} \cos \alpha^{(t)}_{i} + 1}$ is in $(0, 1)$, then $\frac{2}{-(\delta^{(t)}_{i})^{2} + 2\delta^{(t)}_{i} \cos \alpha^{(t)}_{i} + 1}$ is in $(1, 2)$. So based on Lemma~\ref{lem:7:convergence} and Lemma~\ref{lem:7:iteration}, this theorem holds.
% \qed
\end{IEEEproof}

\begin{theorem}
\label{thm:7:privacy}
Algorithm~\ref{alg:7:edpkm} is $\epsilon$-differentially private, where $\epsilon = \epsilon_{0} + \sum_{t = 1}^{\hat{I}}\max_{i = 1}^{k}\{\epsilon^{(t)}_{i}\}$, $\hat{I}$ is its total number of iterations to converge.
\end{theorem}

\begin{IEEEproof}
When applying the ExpDP to sample $\hat{S}^{(t)}_{i}$ (Line 11, Algorithm~\ref{alg:7:edpkm}) in $C^{(t)}_{i}$, we have
\begin{equation*}
\begin{split}
& \frac{\Pr[\hat{S}^{(t)}_{i} = S]}{\Pr[\hat{S}^{\prime(t)}_{i} = S]} \\
= & \frac{\Pr[S,P^{(t)}_{i},S^{(t)}_{i},S^{\prime(t)}_{i}\ in\ a\ plane] \times \exp(\frac{\epsilon^{(t)}_{i}h(\delta^{(t)}_{i},\alpha^{(t)}_{i})}{2\Delta h})}{\Pr[S,P^{(t)}_{i},S^{(t)}_{i},S^{\prime(t)}_{i}\ in\ a\ plane] \times \exp(\frac{\epsilon^{(t)}_{i}h(\delta^{\prime(t)}_{i},\alpha^{\prime(t)}_{i})}{2\Delta h})}\\
\leq & \exp(\epsilon^{(t)}_{i}).
\end{split}
\end{equation*}
So Algorithm~\ref{alg:7:edpkm} guarantees $\epsilon^{(t)}_{i}$-DP in $C^{(t)}_{i}$. Because in each iteration, all the items $x_{i} \in X$ are split into $k$ mutually exclusive clusters, based on the parallel composition and the sequential composition~\cite{MCSHERRY2009}, after $\hat{I}$ iterations, Algorithm~\ref{alg:7:edpkm} is $\epsilon$-differentially private, where $\epsilon = \epsilon_{0} + \sum_{t = 1}^{\hat{I}}\max_{i = 1}^{k}\{\epsilon^{(t)}_{i}\}$. Note that, since the Lloyd's $k$-means algorithm usually converges in small iteration, according to Theorem~\ref{thm:7:iteration} and Theorem~\ref{thm:7:iteration2}, the value of the overall $\epsilon$ would not be very large in expected case.
% \qed
\end{IEEEproof}

% experiments
\section{Experimental Evaluation}
\label{sec:7:experiments}

\subsection{Datasets and Configuration}
Table~\ref{tab:7:dataset} illustrates the key features of the real-world datasets we used to evaluate the clustering quality and the convergence rate of Algorithm~\ref{alg:7:edpkm}. As a matrix, each dataset contains \#Records $\times$ \#Dims cells. We use these datasets with two reasons. Firstly, they are used for the clustering experiments in several research papers for $k$-means clustering tasks, e.g., \cite{KanungoT2002,RodriguezA2014} for normal $k$-means clustering, \cite{ZHANGJ2013,SU2017} for differentially private $k$-means. Secondly, their sizes are in different orders of magnitude, which help us to show the performance stability and the scalability of an algorithm over different datasets.
\begin{table}[!th]
\centering
\caption{Descriptions of Datasets.}
\label{tab:7:dataset}
\scalebox{1}{
\begin{tabular}{|c|c|c|c|}
\hline
Dataset & \#Records & \#Dims & \#Clusters \\ \hline
Iris~\cite{DHEERU2017}    & 150      & 4      & 3          \\ \hline
House~\cite{IMAGE2018}   & 1837     & 3      & 3          \\ \hline
S1~\cite{SSET2006}      & 5000     & 2      & 15         \\ \hline
Birch2~\cite{BIRCH1997}  & 12000    & 2      & 5          \\ \hline
Image~\cite{IMAGE2018}   & 34112    & 3      & 3          \\ \hline
Lifesci~\cite{KOMAREK2018} & 26733    & 10     & 3          \\ \hline
\end{tabular}
}
\end{table}

We compare the clustering quality of Algorithm~\ref{alg:7:edpkm} (in both posterior and prior \textit{sampling zone}) with that of the state-of-the-art $\epsilon$-differentially private $k$-means clustering algorithms and the non-private Lloyd's algorithm. The clustering quality is measured by the difference/gap of the final cost (Equation~\ref{eq:7:kmeans}) between a differentially private $k$-means clustering algorithm and Lloyd's algorithm. A smaller gap indicates better clustering quality. In the experiments, we implement and name them as Posterior (Algorithm~\ref{alg:7:edpkm} with past knowledge), Prior (Algorithm~\ref{alg:7:edpkm} with future knowledge \cite{LuZ2019}), SU~\cite{SU2017}, PrivGene~\cite{ZHANGJ2013}, GUPT~\cite{MOHAN2012}, DWORK~\cite{DWORK2011}, BLUM~\cite{BLUM2005}, and LLOYD~\cite{LLOYD1982}.

Because the six algorithms achieve $\epsilon$-DP are randomised, we report their expected clustering quality. According to the law of large numbers, we run all the seven algorithms 300 times and take the average results as the expectations. The initial set of centroids is randomly selected for all methods in each run. For those relying on a predefined iteration number, we take the corresponding value (or function) from the original papers. In addition, we normalise the data in all the datasets to $[0,1]$. Furthermore, we normalise the final cost for all involved algorithms, i.e., the final cost of Lloyd's algorithm is always one. 

In addition, in each run, LLOYD, BLUM, DWORK, SU, and Algorithm~\ref{alg:7:edpkm} use the same initial centroids. Because GUPT starts from splitting the original datasets into several buckets, it cannot use the same initial centroids as LLOYD. Note that calculating the overall privacy budget depends on whether a method converges. Algorithm~\ref{alg:7:edpkm} and GUPT calculate the overall privacy budget bottom-up. That is, once it terminates, we sum all the privacy budgets used in each iteration to have the overall privacy budget. SU, PrivGene, DWORK, and BLUM calculate it top-down. Namely, the given overall privacy budget is split to each iteration at the initialisation step. Therefore, in the experiments, we first allocate the same privacy budget to each atom step for Algorithm~\ref{alg:7:edpkm} and GUPT, then calculate their overall privacy budgets. Next we take the overall privacy budget of Algorithm~\ref{alg:7:edpkm} as the overall privacy budget for the methods cannot converge. In the experiments, local sensitivity is applied for all DP algorithms.

\subsection{Experimental Results}
\begin{figure*}[!th]
     \subfloat[Iris ($k$ = 3)]{%
       \includegraphics[width=0.33\textwidth]{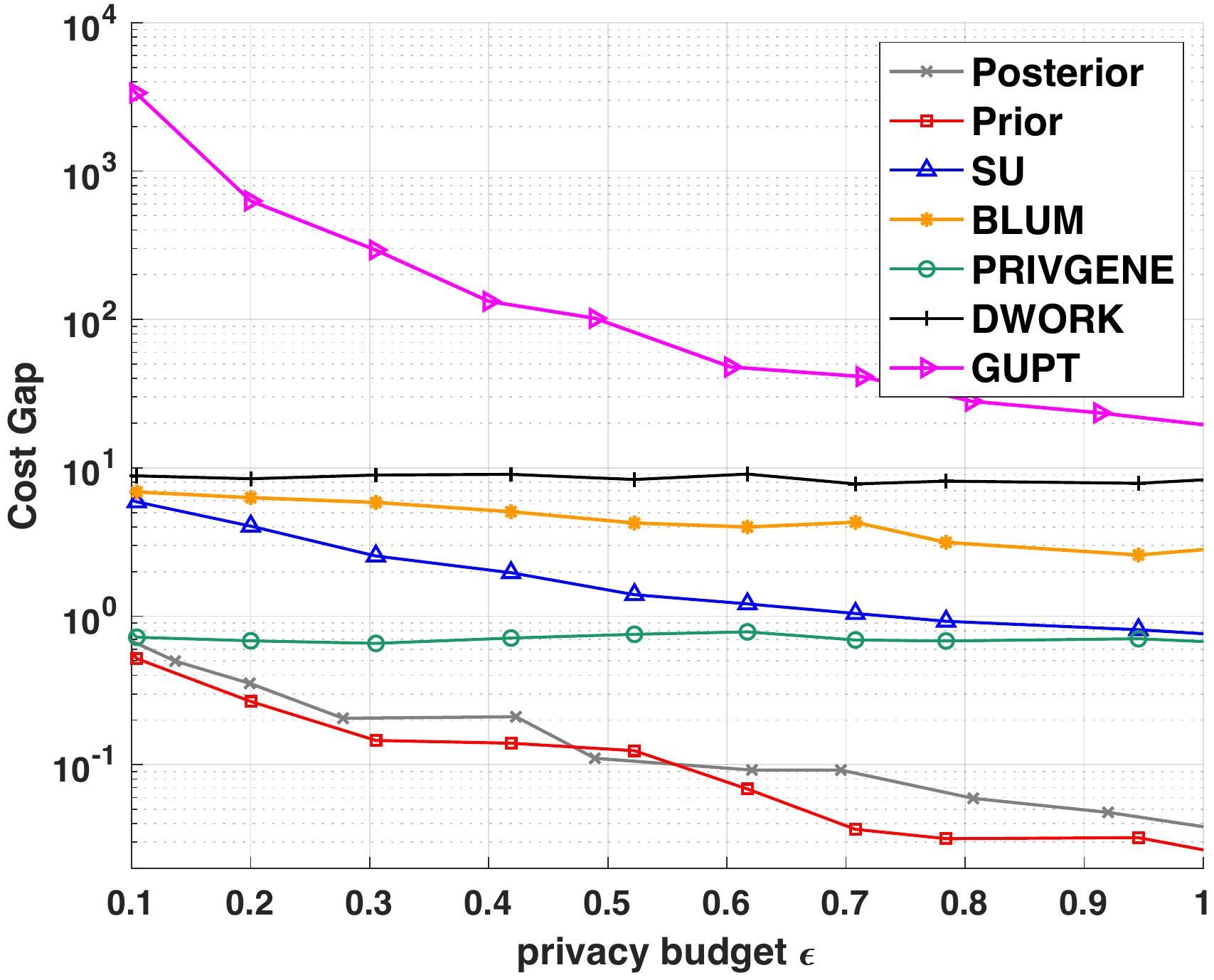}
       \label{subfig:7:iris}
     }
     \hfill
     \subfloat[House ($k$ = 3)]{%
       \includegraphics[width=0.32\textwidth]{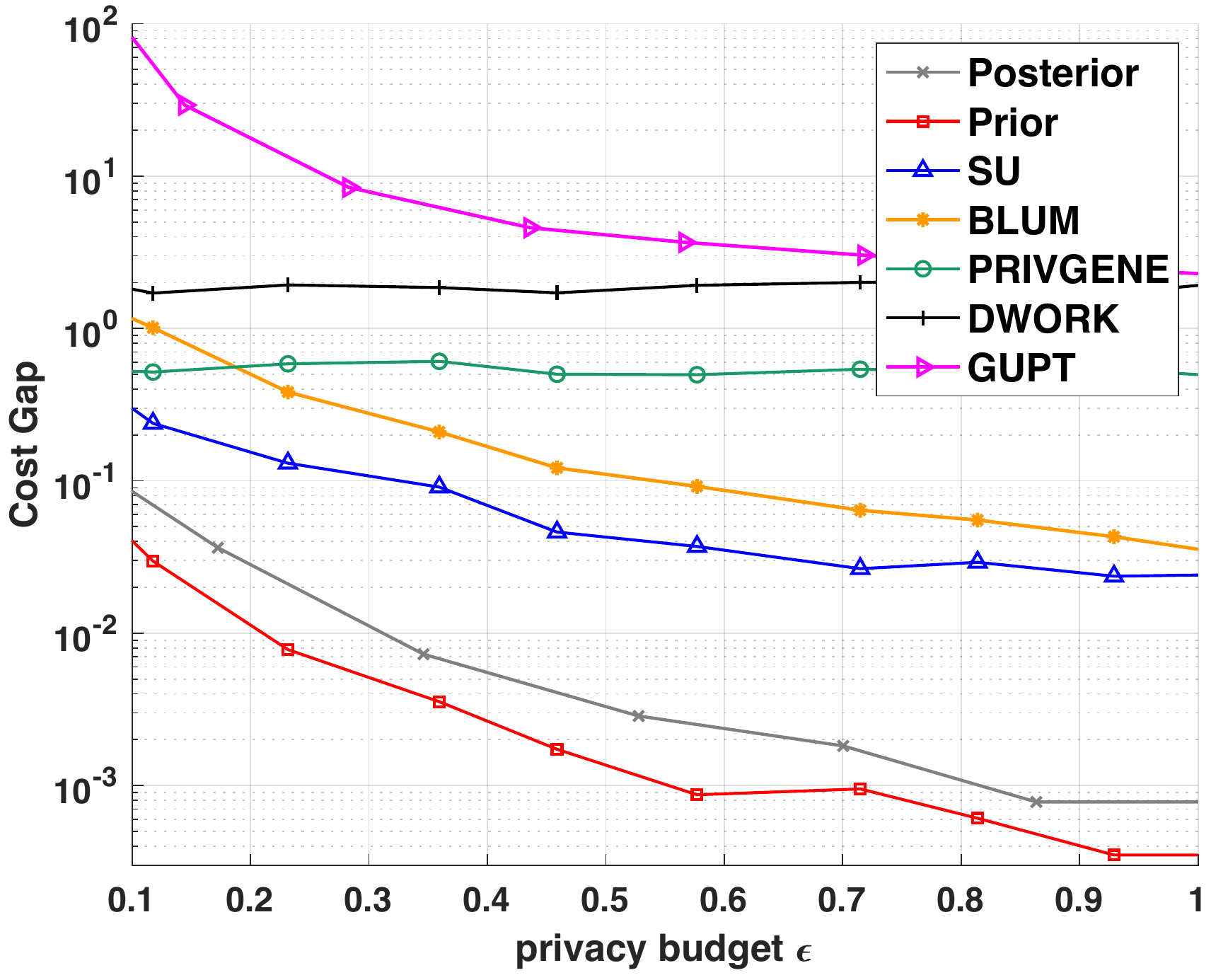}
       \label{subfig:7:house}
     }
     \hfill
     \subfloat[S1 ($k$ = 15)]{%
       \includegraphics[width=0.32\textwidth]{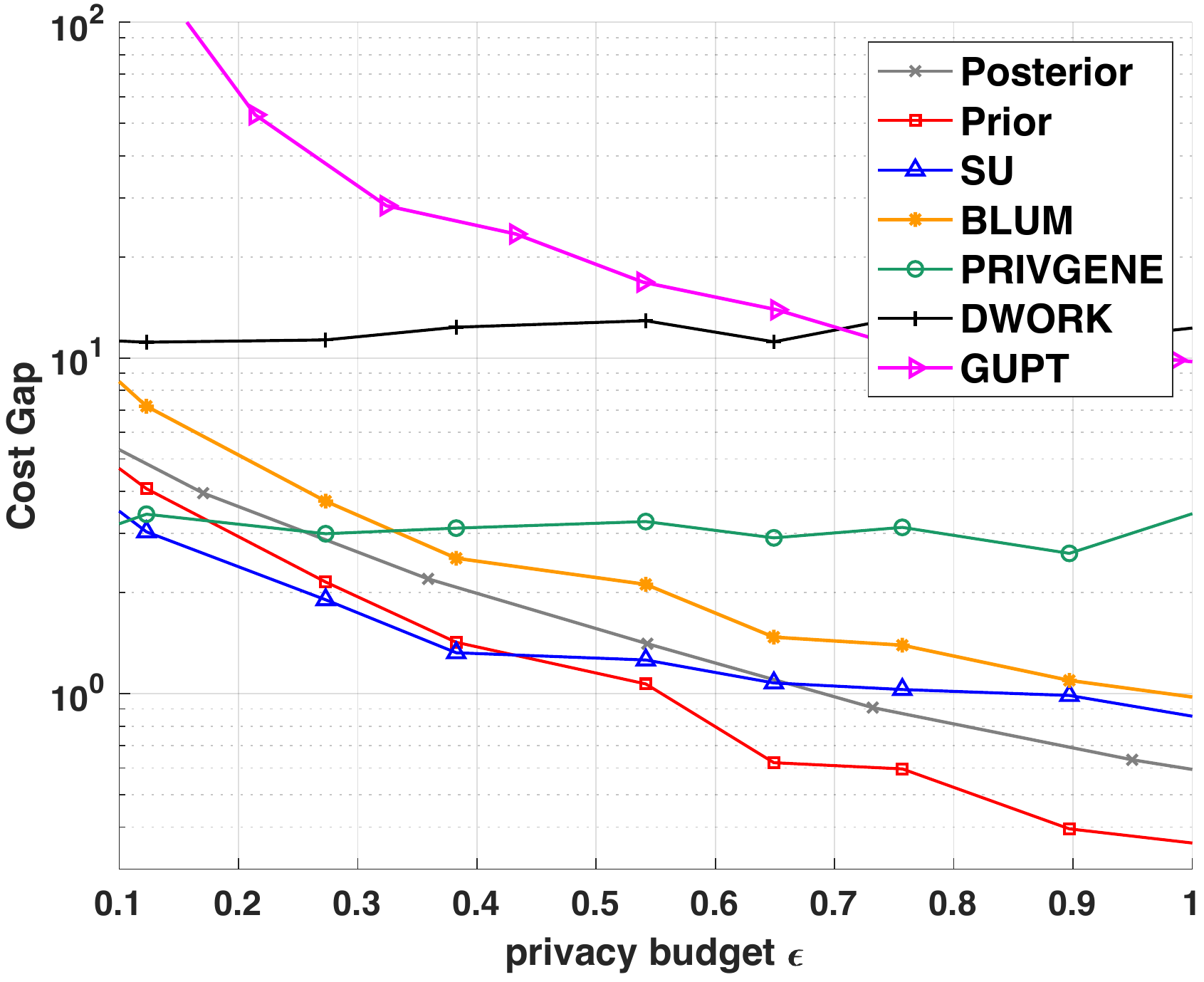}
       \label{subfig:7:s1}
     }
     \hfill
     \subfloat[Birch2 ($k$ = 5)]{%
       \includegraphics[width=0.32\textwidth]{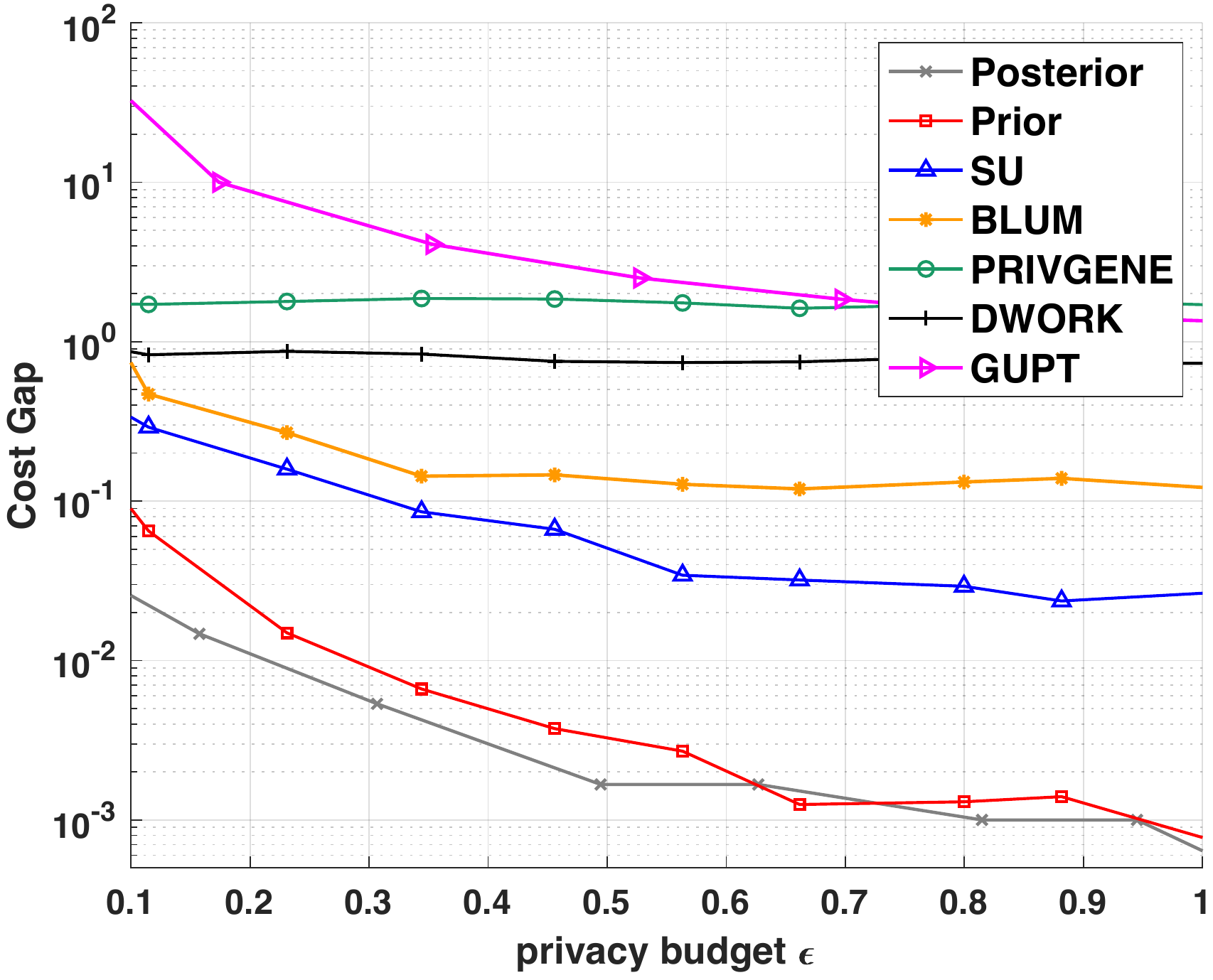}
       \label{subfig:7:birch2}
     }
     \hfill
     \subfloat[Image ($k$ = 3)]{%
       \includegraphics[width=0.32\textwidth]{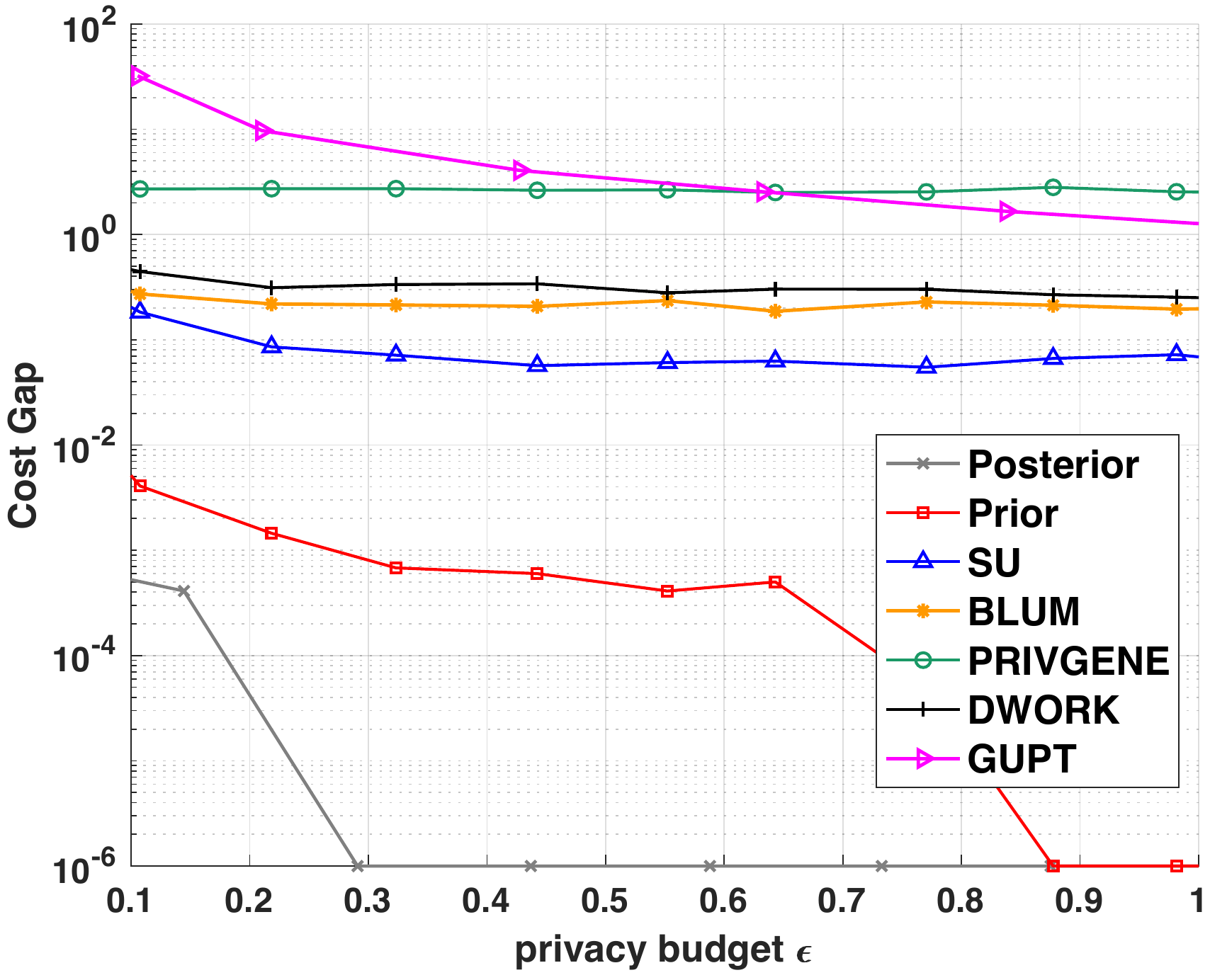}
       \label{subfig:7:image}
     }
     \hfill
     \subfloat[Lifesci ($k$ = 3)]{%
       \includegraphics[width=0.32\textwidth]{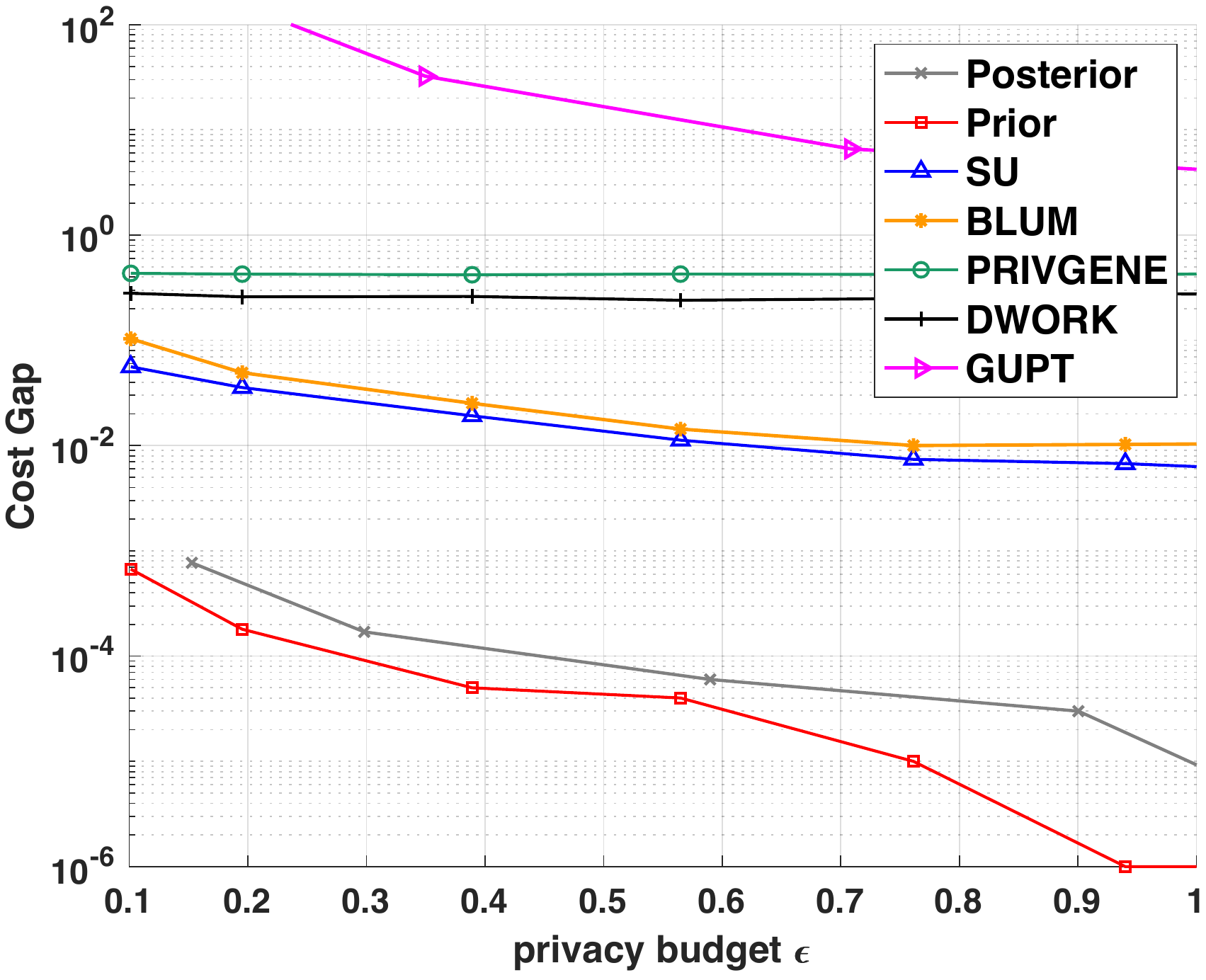}
       \label{subfig:7:lifesci}
     }
     \caption{Clustering Quality Comparisons.}
     \label{fig:7:clustering}
\end{figure*}

% comparison
Figure~\ref{fig:7:clustering} reports the expected clustering quality of each algorithm, where the cost gap is in $\log$ scale, the privacy budget is varied in $[0.1, 1.0]$. Generally, Algorithm~\ref{alg:7:edpkm} outperforms the state-of-the-art results with the same DP requirement in the six datasets in both posterior and prior cases. Additionally, the performance gap between Algorithm~\ref{alg:7:edpkm} and the existing algorithms increases when increasing $\epsilon$, which indicates better trade-off between privacy and utility with our algorithm. Furthermore, Algorithm~\ref{alg:7:edpkm} performs much better than other algorithms in the larger datasets (e.g., Image and Lifesci), which reflects the potentially good scalability of our algorithm.

\begin{figure*}[!th]
     \subfloat[Convergence Degree of the Posterior Strategy (Past Knowledge)]{%
       \includegraphics[width=0.45\textwidth]{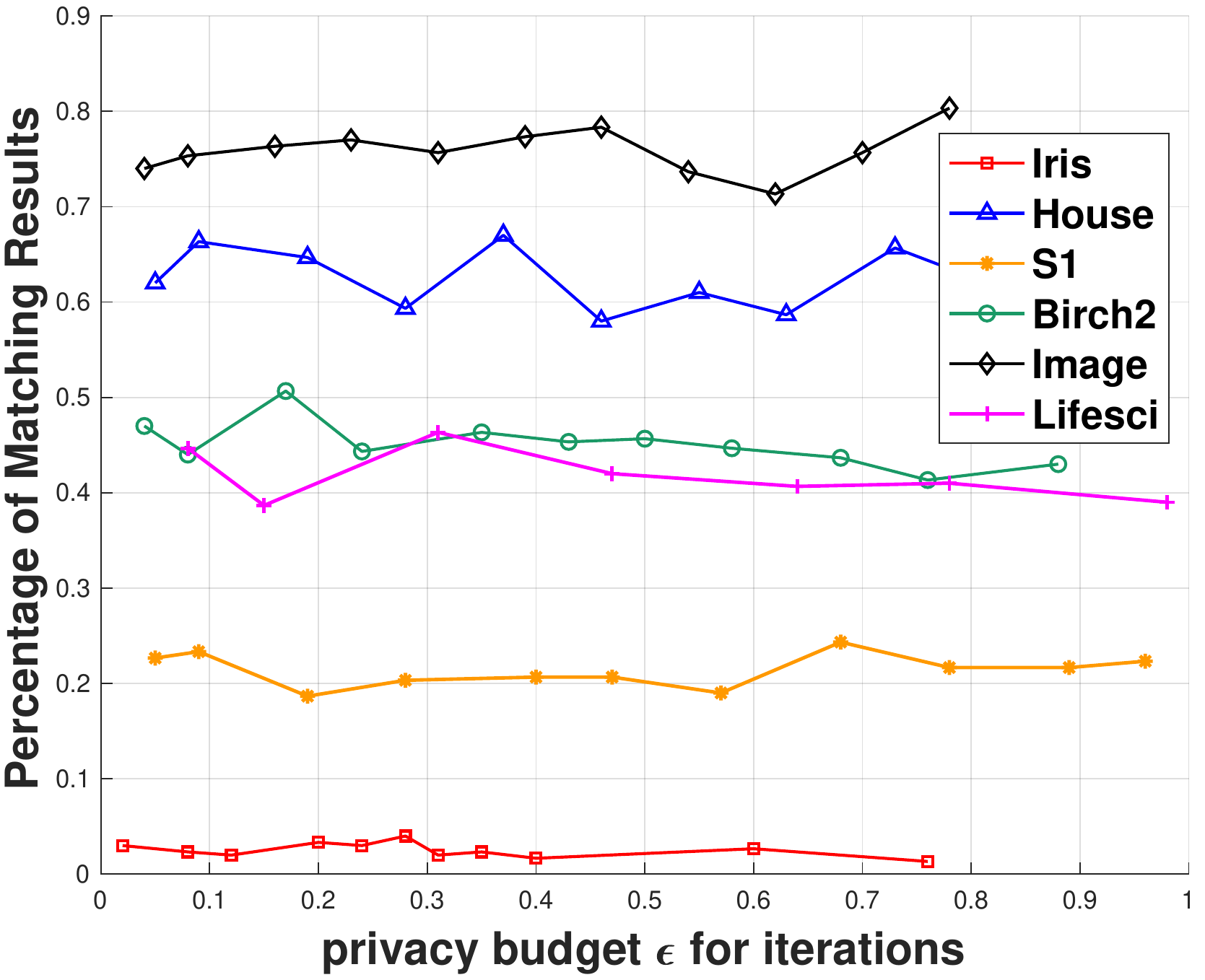}
       \label{subfig:7:perc_post}
     }
     \hfill
     \subfloat[Convergence Degree of the Prior Strategy (Past and Future Knowledge)]{%
       \includegraphics[width=0.45\textwidth]{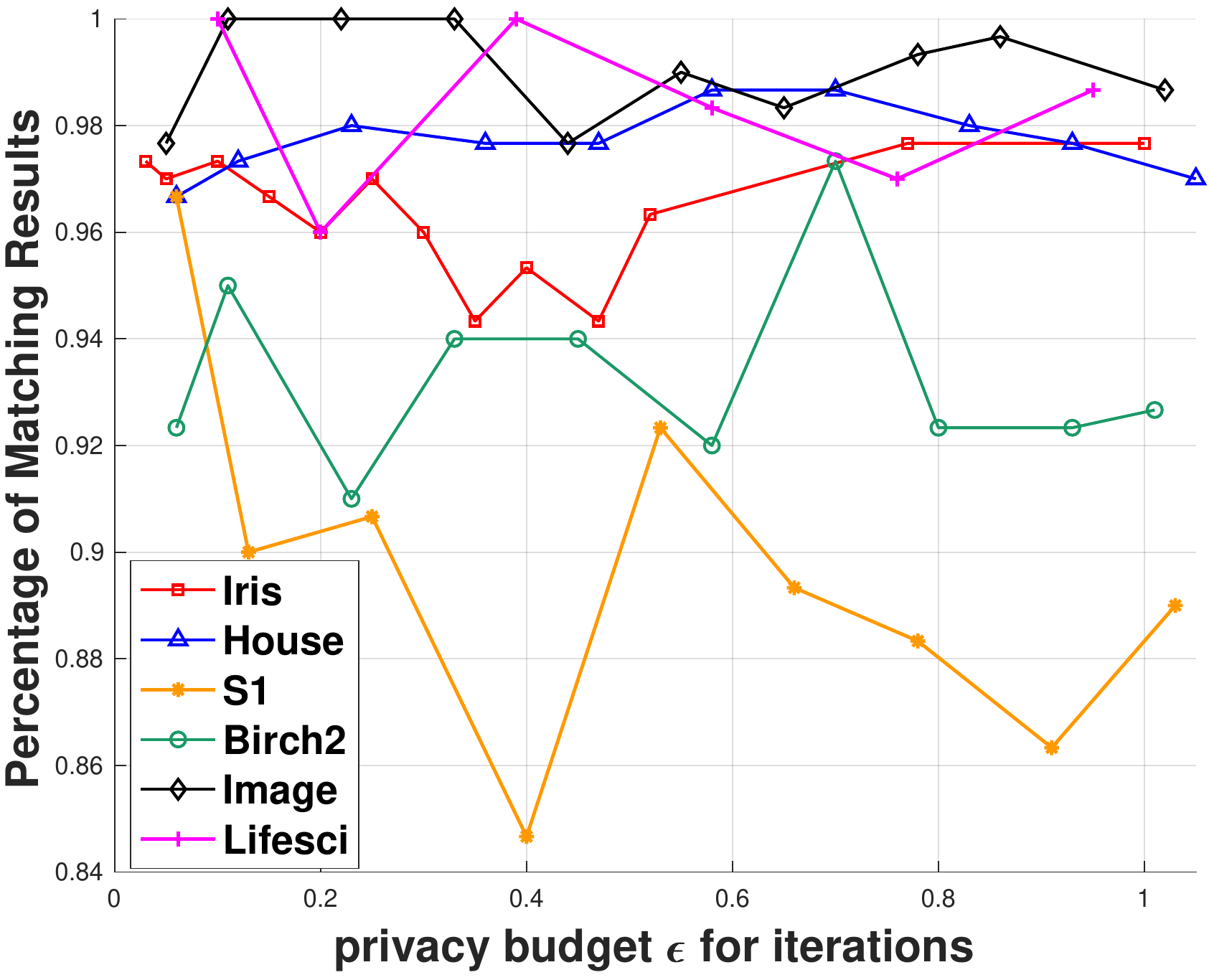}
       \label{subfig:7:perc_prior}
     }
     \caption{Convergence Degree of Our Approach.}
     \label{fig:7:convergence}
\end{figure*}

% convergence
Figure~\ref{fig:7:convergence} depicts the convergence degree of Algorithm~\ref{alg:7:edpkm} (in both two strategies). We study the convergence degree by comparing whether the output set of centroids of our approach (without the final DP noise as Line 12 in Algorithm~\ref{alg:7:edpkm}) is the same to that of Lloyd's algorithm. Since we round the values in the clustering process, once the output of our approach is in $[0.99, 1.01]$ of Lloyd's algorithm, we call it a match in this paper. We report the percentage of the matching results of the two strategies over all six datasets as the convergence degree. From Figure~\ref{fig:7:convergence}, the prior strategy, which uses both past and future knowledge, outperforms the posterior strategy, which only uses the past knowledge, in the convergence performance because of the convergence guarantee of Theorem~\ref{thm:7:sameconv}. Particularly, the prior strategy matches at least 84\% (90\% in most cases) output centroids of Lloyd's algorithm; while the posterior matches no more than 80\% (50 \% in most cases).

\begin{figure*}[!th]
  \centering
  \subfloat[Iteration Ratio over $\epsilon$.]{
    \adjustbox{raise=-1.8pc}{
      \includegraphics[height=5cm]{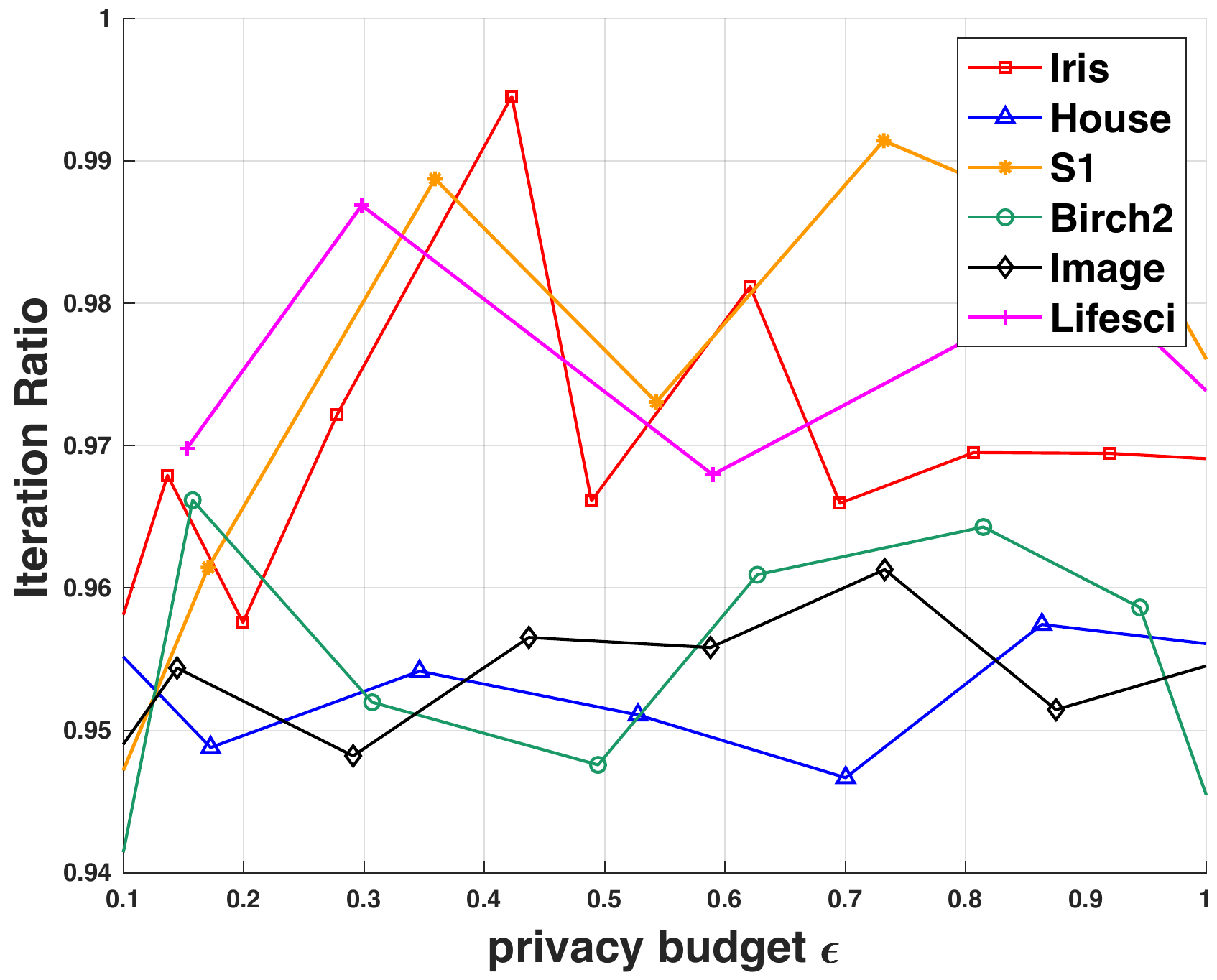}
    }
  }
  \subfloat[Average Iterations to Convergence.]{%
    \scalebox{1}{
      \begin{tabular}{|c|c|c|c|c|c|c|}
      \hline
          & Iris & House & S1   & Birch2 & Image & Lifesci \\ \hline
      Posterior & 5.89  & 16.40  & 17.05 & 14.73   & 13.60  & 28.92    \\ \hline
      LLOYD    & 6.07  & 17.22  & 17.64 & 15.52   & 14.26  & 29.72    \\ \hline
      Ratio    & 0.97  & 0.95   & 0.97  & 0.95    & 0.95   & 0.97     \\ \hline
      \end{tabular}
    }
  }
  \caption{Iterations of the Posterior Strategy (Past Knowledge) to Convergence.}
  \label{fig:7:iter_posterior}
\end{figure*}

\begin{figure*}[!th]
  \centering
  \subfloat[Iteration Ratio over $\epsilon$.]{
    \adjustbox{raise=-1.8pc}{
      \includegraphics[height=5cm]{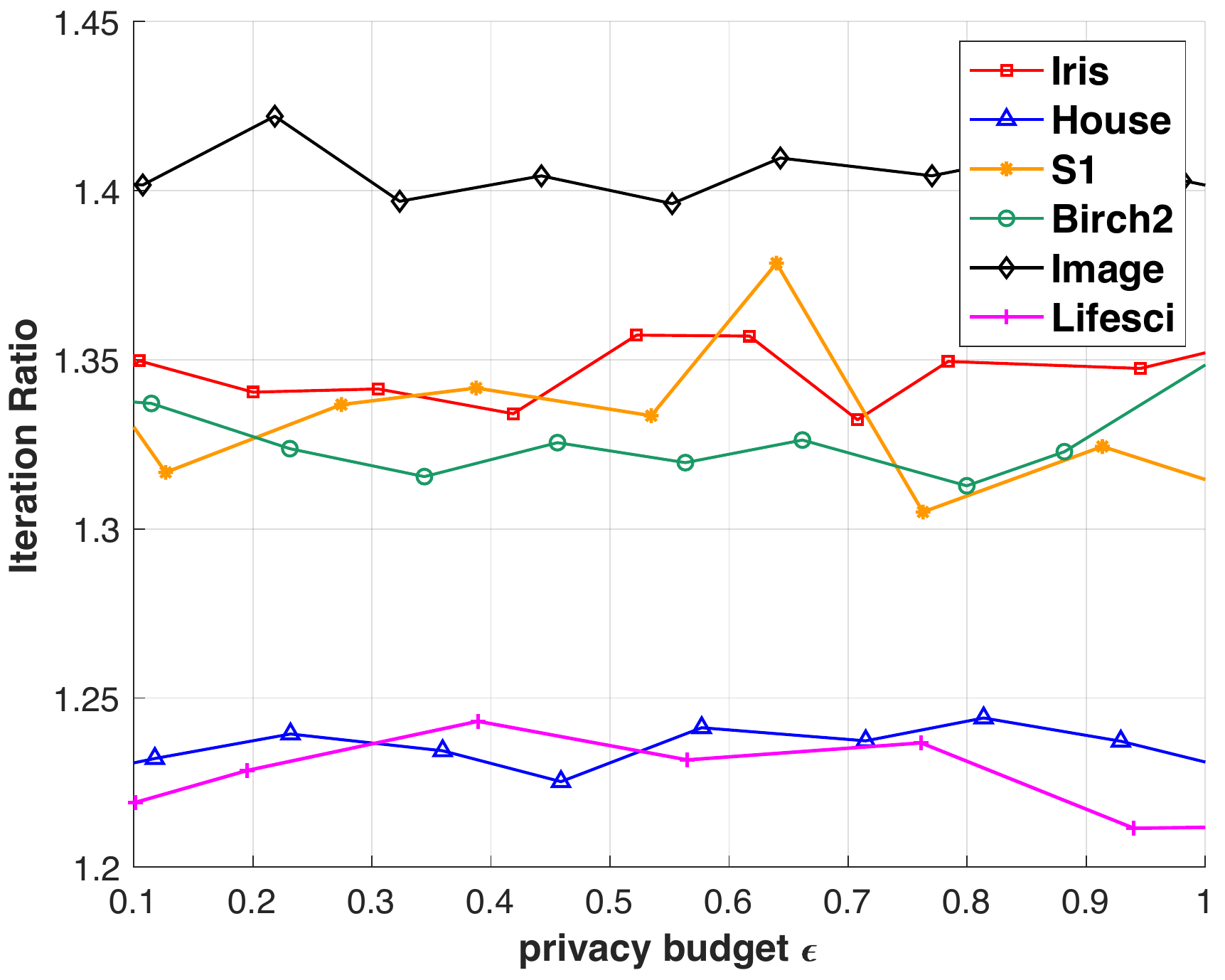}
    }
  }
  \subfloat[Average Iterations to Convergence.]{%
    \scalebox{1}{
      \begin{tabular}{|c|c|c|c|c|c|c|}
      \hline
          & Iris & House & S1   & Birch2 & Image & Lifesci \\ \hline
      Prior & 8.2  & 21.4  & 23.9 & 20.7   & 19.8  & 36.6    \\ \hline
      LLOYD    & 6.1  & 17.3  & 18.0 & 15.6   & 14.1  & 29.9    \\ \hline
      Ratio    & 1.34  & 1.24   & 1.33  & 1.33    & 1.40   & 1.22     \\ \hline
      \end{tabular}
    }
  }
  \caption{Iterations of the Prior Strategy (Past and Future Knowledge) to Convergence.}
  \label{fig:7:iter_prior}
\end{figure*}

% iteration
Figure~\ref{fig:7:iter_posterior} and Figure~\ref{fig:7:iter_prior} show the iteration ratio between Algorithm~\ref{alg:7:edpkm} and Lloyd's algorithm to converge, which confirms the theoretical analysis in Theorem~\ref{thm:7:iteration2} and Theorem~\ref{thm:7:iteration}. In particular, we compare the numbers of iterations  that  Algorithm~\ref{alg:7:edpkm} and Lloyd's algorithm execute till termination. Note that, in the experiments, the privacy budget does not impacts the number of iterations significantly because the experimental performance of the ExpDP is not as good as its theoretical guarantee with a relatively small \textit{sampling zone}.

\section{Conclusion}
\label{sec:7:conclusion}
To address the non-convergence problem in the existing algorithms for differentially private $k$-means clustering in the interactive setting, in this paper, we proposed a novel centroids updating approach by applying the exponential mechanism of differential privacy in a selected area. The novelty of our approach is the orientation control of centroid movement for noise injection in the iterations of the clustering process to achieve convergence. We proved the key properties of our approach and showed that it converges in at most twice as many iterations as Lloyd's $k$-means algorithm. The experimental evaluations validated that with the same DP guarantee, our algorithm ensures convergence and achieves better clustering quality than the state-of-the-art differentially private algorithms in the interactive setting.

% if have a single appendix:
\appendix[Proof of Equation~\ref{eq:7:gap} in the Proof of Lemma~\ref{lem:7:convergence}]
\label{sec:7:apdxa}
Assume each record $x_i$ in the dataset $X$ has $d$ dimensions. We will have:
$$J^{(S^{(t - 1)}_{i})} = \sum_{j = 1}^{||C^{(t)}_{i}||} ||x_{j} - S^{(t - 1)}_{i}||^{2} = \sum_{j = 1}^{||C^{(t)}_{i}||}\sum_{p = 1}^{d} \left(x_{jp} - S^{(t - 1)}_{ip}\right)^{2},$$
$$J^{(S^{(t)}_{i})} = \sum_{j = 1}^{||C^{(t)}_{i}||} ||x_{j} - S^{(t)}_{i}||^{2} = \sum_{j = 1}^{||C^{(t)}_{i}||}\sum_{p = 1}^{d} \left(x_{jp} - S^{(t)}_{ip}\right)^{2}.$$
The distance $a^{(t)}_{i}$ between $S^{(t - 1)}_{i}$ and $S^{(t)}_{i}$ can be further split to $(a^{(t)}_{i})^{2} = \sum_{p = 1}^{d} (a^{(t)}_{ip})^{2}$, where $a^{(t)}_{ip} = S^{(t - 1)}_{ip} - S^{(t)}_{ip}$. Note that $a^{(t)}_{i} > 0$, $a^{(t)}_{ip}$ can be any real number. Then
\begin{equation*}
\begin{split}
& J^{(S^{(t - 1)}_{i})} - J^{(S^{(t)}_{i})} \\
= & \sum_{j = 1}^{||C^{(t)}_{i}||}\sum_{p = 1}^{d} \left[(x_{jp} - S^{(t - 1)}_{ip})^{2} - (x_{jp} - S^{(t)}_{ip})^{2}\right] \\
= & \sum_{j = 1}^{||C^{(t)}_{i}||}\sum_{p = 1}^{d} \left[(S^{(t - 1)}_{ip} - S^{(t)}_{ip})(S^{(t - 1)}_{ip} + S^{(t)}_{ip} - 2x_{jp}\right] \\
= & \sum_{j = 1}^{||C^{(t)}_{i}||}\sum_{p = 1}^{d} \left[(S^{(t - 1)}_{ip} - S^{(t)}_{ip})(S^{(t - 1)}_{ip} - S^{(t)}_{ip} + 2S^{(t)}_{ip} - 2x_{jp})\right] \\
= & \sum_{j = 1}^{||C^{(t)}_{i}||}\sum_{p = 1}^{d} \left\{a^{(t)}_{ip}[a^{(t)}_{ip} - 2(x_{jp} - S^{(t)}_{ip})]\right\} \\
= & \sum_{j = 1}^{||C^{(t)}_{i}||}\sum_{p = 1}^{d} \left[(a^{(t)}_{ip})^{2} - 2a^{(t)}_{ip}(x_{jp} - S^{(t)}_{ip})\right] \\
= & \sum_{j = 1}^{||C^{(t)}_{i}||}\sum_{p = 1}^{d} \left(a^{(t)}_{ip}\right)^{2} - 2\sum_{p = 1}^{d} \left[a^{(t)}_{ip} \times \sum_{j = 1}^{||C^{(t)}_{i}||}(x_{jp} - S^{(t)}_{ip})\right] \\
= & ||C^{(t)}_{i}|| \times (a^{(t)}_{i})^{2} - 2\sum_{p = 1}^{d} \left(a^{(t)}_{ip} \times 0\right) \\
= & ||C^{(t)}_{i}|| \times (a^{(t)}_{i})^{2}
\end{split}
\end{equation*}
% or
%\appendix  % for no appendix heading
% do not use \section anymore after \appendix, only \section*
% is possibly needed

% use appendices with more than one appendix
% then use \section to start each appendix
% you must declare a \section before using any
% \subsection or using \label (\appendices by itself
% starts a section numbered zero.)
%

% \appendices
% \section{Proof of the First Zonklar Equation}
% Appendix one text goes here.

% % you can choose not to have a title for an appendix
% % if you want by leaving the argument blank
% \section{}
% Appendix two text goes here.

% use section* for acknowledgment
\ifCLASSOPTIONcompsoc
  % The Computer Society usually uses the plural form
  \section*{Acknowledgments}
\else
  % regular IEEE prefers the singular form
  \section*{Acknowledgment}
\fi

This work is supported by Australian Government Research Training Program Scholarship, Australian Research Council Discovery Project DP150104871, National Key R \& D Program of China Project \#2017YFB0203201, and supported with supercomputing resources provided by the Phoenix HPC service at the University of Adelaide. The corresponding author is Hong Shen.

% Can use something like this to put references on a page
% by themselves when using endfloat and the captionsoff option.
% \ifCLASSOPTIONcaptionsoff
%   \newpage
% \fi

% trigger a \newpage just before the given reference
% number - used to balance the columns on the last page
% adjust value as needed - may need to be readjusted if
% the document is modified later
%\IEEEtriggeratref{8}
% The "triggered" command can be changed if desired:
%\IEEEtriggercmd{\enlargethispage{-5in}}

% references section

% can use a bibliography generated by BibTeX as a .bbl file
% BibTeX documentation can be easily obtained at:
% http://mirror.ctan.org/biblio/bibtex/contrib/doc/
% The IEEEtran BibTeX style support page is at:
% http://www.michaelshell.org/tex/ieeetran/bibtex/
\bibliographystyle{IEEEtran}
% argument is your BibTeX string definitions and bibliography database(s)
\bibliography{zglu}
\end{document}